\documentclass[pra,a4paper,twocolumn,showpacs,superscriptaddress]{revtex4-1}

\usepackage{epsf,epsfig}
\usepackage[psamsfonts]{amssymb}
\usepackage{amsmath}
\usepackage{bm}
\usepackage{natbib}

\def\imath{\rm i}
\def\G{\mathord{\buildrel{\lower3pt\hbox{$\scriptscriptstyle\leftrightarrow$}}
\over G}}

\newcommand{\bfsfG}{\mbox{\sffamily\bfseries{G}}}

\begin{document}
\title{Casimir forces in multilayer magnetodielectrics with both gain and loss}
\author{Ehsan Amooghorban}
\email{amooghorban@sci.ui.ac.ir}
\affiliation{Department of Photonics Engineering, Technical University of Denmark, DK-2800 Kgs. Lyngby, Denmark}
\affiliation{Department of Physics, University of Isfahan, Hezar Jarib Ave., Isfahan, Iran}

\author{Martijn Wubs}
\affiliation{Department of Photonics Engineering, Technical University of Denmark, DK-2800 Kgs. Lyngby, Denmark}

\author{N. Asger Mortensen}
\affiliation{Department of Photonics Engineering, Technical University of Denmark, DK-2800 Kgs. Lyngby, Denmark}

\author{Fardin Kheirandish}
\affiliation{Department of Physics, University of Isfahan, Hezar Jarib Ave., Isfahan, Iran}

\begin{abstract}
\noindent A path-integral approach to the quantization of the
electromagnetic field in a linearly amplifying
magnetodielectric medium is presented. Two continua of inverted
harmonic oscillators are used to describe the polarizability and
magnetizability of the amplifying medium. The causal
susceptibilities of the amplifying medium, with negative imaginary
parts in finite frequency intervals, are identified and their
relation to microscopic coupling functions are determined. By
carefully relating the two-point functions of the field theory to
the optical Green functions, we calculate the Casimir energy and
Casimir forces for a multilayer magnetodielectric medium with both
gain and loss. We point out the essential differences with a purely
passive layered medium. For a single layer, we find different bounds
on the Casimir force for fully amplifying and for lossy media.
The force is attractive in both cases, also if the medium exhibits
negative refraction. From our Lagrangian we also derive by canonical
quantization the postulates of the phenomenological theory of amplifying
magnetodielectrics.
\end{abstract}



\date{\today}


\maketitle


\section{Introduction}\label{Sec:Intro}
The Casimir force is a pure quantum effect that can be considered as
the macroscopic manifestation of the vacuum fluctuations of the
electromagnetic fields in the presence of boundaries. Originally
derived in 1948 for two ideally conducting or reflecting plates in
vacuum~\cite{Casimir1948a},
 the Casimir force per area $F_{\rm C}/A$ was found to be $- \hbar c \pi^{2}/(240 d^4)$,
an attractive force with characteristic inverse fourth power dependence on the plate separation $d$.
A theory for Casimir forces between parallel dielectrics was
developed by Lifshitz {\em et al.}~\cite{Lischitz1956a}, further
refined by Schwinger {\em et al.}~\cite{Schwinger1978a}, and since
then extended to arbitrary multilayer
dielectrics~\cite{Zhou1995a,Klimchitskaya2000a,Tomas2002a,Raabe2003a,Ellingsen2007a}
and other geometries.

In recent years, the Casimir force has become technologically relevant,
with the development of micro- and nano-electromechanical
systems with small components at close proximity of each other.
 On the one hand, the Casimir force can be a major cause of
stiction ({\em i.e.} microscopic components sticking together),
friction, or adhesion, and thus forms a possible obstacle for the
operation of nanostructured devices. On the other hand, a novel
class of  microelectromechanical systems (MEMS) only works because
of the Casimir force~\cite{Chan2001a,Chan2001b}.

It is therefore both fundamentally interesting and technologically
relevant to what extent the Casimir force can be controlled by
changing the electromagnetic environment. For most geometries, the
Casimir force between two media separated by vacuum is an attractive
force,  with a magnitude that becomes appreciable in the submicron
range and rapidly increases in the nanometer range. However, already
Lifschitz predicted that the Casimir force for parallel dielecric
layers can be attractive or repulsive, depending on the relative
values of the dielectric constants of the successive
layers~\cite{Lischitz1956a}. The first experimental confirmation of
his prediction came only recently: Munday {\em et al.} found Casimir
forces with different signs for suitably chosen interacting
materials immersed in a fluid~\cite{Munday2009a}, with a measured
repulsive interaction which was weaker than the attractive
counterpart. Further
 measurements of Casimir forces are reported
in~\cite{Spaarnaay1958a,Sabisky1973a,Lamoreaux1998a,Mohideen1998a,
Chan2008a,Hertlein2008a,Kim2009a,Man2009a,Masuda2009a}.

If Casimir forces could be made repulsive, then this could eliminate
the unwanted phenomenon of stiction~\cite{Butt2005a}, enable quantum
levitation of objects in a fluid, and lead to new classes of
switchable nanoscale devices with ultra-low static
friction~\cite{Capasso2007a,Iannuzzi2005a,Feiler2008a,Rosa2008a}. So
what options exist to make the Casimir force repulsive, besides
Lifschitz's suggestion?
In the first place, metamaterials have been proposed to this end.
However, loss in metallic substructures may intuitively turn a
repulsive force into an attractive one, which may explain why
repulsive Casimir forces with metamaterials have not yet been
reported~\cite{Lamoreaux2009a,Pirozhenko2008a}. Recently, Zhao {\em
et al.} showed theoretically that a repulsive Casimir force could be
realized with
 metamaterials with strong chirality~\cite{Zhao2010a}.
Another mechanism to obtain  repulsive Casimir forces is Boyer's
Casimir repulsion based on an asymmetric three-layer setup of a
nonmagnetic medium on the one end and a purely magnetic medium on the other,
separated by vacuum~\cite{Boyer1974a,Kenneth2002a}. It relies on the nontrivial
possibility of developing new artificial negative-index
metamaterials~\cite{Pirozhenko2008a}.

A third option to obtain repulsive Casimir forces, and
the one considered in this paper, is the use of media with optical
gain~\cite{Sherkunov2005a,Leonhardt2007a}.
Indeed, the main aim of the present article is to determine Casimir energy
and forces in amplifying magnetodielectrics.
With amplifying medium, we mean a medium for which the imaginary part of the electric or the magnetic susceptibility becomes negative for one or more frequency intervals (${\rm Im}[\varepsilon(\omega)]\equiv \varepsilon_{\rm I}(\omega)
<0$ or $\mu_{\rm I}(\omega) <0$, in contrast to lossy systems for which both $\varepsilon_{\rm I}(\omega)$ and $\mu_{\rm I}(\omega)$ are always positive.
We allow for gain not only in the electric but also in the magnetic
response, thereby treating electric and magnetic fields on an equal
footing in our theory.

Since it is a quantum mechanical effect, the calculation of the
Casimir force for media with gain requires a consistent procedure
for quantization of the electromagnetic field in presence of amplifying
medium. This can be done with the concept of inverted quantum
harmonic oscillators that was introduced by Glauber~\cite{Glauber1986a}. Generally, a rigorous quantization
procedure would require a Lagrangian and Hamiltonian formulation of
the theory, followed by the standard canonical quantization rules.
Currently, a consistent phenomenological approach exists to
macroscopic quantum electrodynamics in presence of amplifying media,
but no canonical formulation is
attempted~\cite{Jeffers1993a,Jeffers1996a,Matloob1997a,Scheel1998a,Raabe2008a}.
Below we derive the postulates of this phenomenological theory from
a canonical quantum theory, where continua of inverted oscillators
are used to describe linearly amplifying media.

The Casimir force in the presence of amplifying materials, with or without negative index, has only recently begun to be explored.
Leonhardt and Philbin calculated the effects of an amplifying
dielectric on the Casimir force, based on the assumption that the
well-known Lifshitz formula for the Casimir force is applicable
without change to amplifying media~\cite{Leonhardt2007a}, and found that
 the Casimir force in the presence of this
medium is repulsive. We will address this issue here as well.

 Sambale {\em
et al.}~\cite{Sambale2009a,Sambale2010a} apply the phenomenological
quantization of the electromagnetic field for amplifying
magnetodielectric media to calculate the Casimir and
Casimir--Polder forces. They find  that the Casimir--Polder force on a weakly
polarisable plate of excited gas atoms is attractive at short distances from a mirror,
and oscillating behavior between attraction and repulsion for larger plate-mirror separations.
However, at a more technical
level there is some controversy about the applicability of the
Minkowski stress tensor and an alternative, Lorentz force-based
tensor was proposed~\cite{Welsch2005a}. This approach was  adopted in Refs.~\cite{Sambale2009a,Sambale2010a} to calculate the Casimir(--Polder) forces, but disputed by
Pitaevskii~\cite{Pitaevskii2006a} and Brevik~\cite{Brevik2009a}.

Our approach to calculate Casimir forces is different. We first develop a
path-integral method for the quantization of the electromagnetic
field in linearly amplifying magnetodielectrics. We benefit from and
generalize recent results obtained with path integrals. In
Ref.~\cite{Li1991a}, Li and Kardar developed a path-integral
approach for computing fluctuation-induced forces between manifolds
immersed in a correlated fluid. Golestanian and Kardar extended this
formalism to arbitrary but small deformations of the boundaries and
focused on the mechanical response of the
vacuum~\cite{Golestanian1998a}. Emig and colleagues also used the
path-integral formalism to obtain the normal and lateral Casimir
forces between two sinusoidally corrugated perfectly conducting
surfaces~\cite{Emig2001a}. Recently, one of the authors extended
this formalism to calculate the Casimir force between two perfectly
conducting plates immersed in a magnetodielectric
medium~\cite{Kheirandish2010a}. Here we extend this quantization
scheme to arbitrary multilayer amplifying media.

The structure of the paper is as follows: In Sec.~\ref{Sec:quantization}, we propose a
Lagrangian for the electromagnetic field in an amplifying
magnetodielectric medium and derive the generating function, which is
used in Sec.~\ref{Sec:Greentensors} for the path integral quantization. Causal electric
and magnetic susceptibilities of the amplifying medium are obtained,
both in the frequency interval(s) with gain and in the remaining lossy regions. We calculate the Green tensor and Casimir forces for a
multilayer amplifying magnetodielectric medium in Sec.~\ref{Sec:Casimir}, and corresponding numerical results are presented in Sec.~\ref{Sec:Numerical}. In Sec.~\ref{Sec:canonical} we derive from our Lagrangian a canonical theory that supports the known phenomenological approach to quantization in amplifying magnetodielectrics. We conclude in Sec.~\ref{Sec:Conclusions}. Further
details of our calculations are given in two appendices.


\section{Field quantization}\label{Sec:quantization}
The quantum electrodynamics of a linearly damped magnetodielectric
medium can be described by modeling the medium as two independent
reservoirs that interact with the electromagnetic field. Each
reservoir contains a continuum of three-dimensional harmonic
oscillators that describe the polarizability and magnetizability of
the
medium~\cite{Huttner1992a,Wubs2001a,Suttorp2004a,Kheirandish2006a,Kheirandisch2008a,Kheirandish2010b,Philbin2010a}.
We assume the medium is linearly amplifying, at least in one or more
finite frequency windows where $\varepsilon_{\rm I}(\omega)<0$ and/or
$\mu_{\rm I}(\omega)<0$. Despite the striking practical differences
between amplifying and lossy media, their theoretical descriptions
turn out to be quite similar. We adopt Glauber's inverted oscillator
to model the quantum amplifier~\cite{Glauber1986a} and use  continua
of inverted oscillators to describe gain instead of loss in the
polarizability and magnetizability of the medium.

We introduce our model for optical media with both gain and loss by
first specifying its Lagrangian density in real space,
\begin{equation}\label{total Lagrangian}
{\cal L}={\cal L}_{\rm EM}+{\cal L}_{\rm e}+{\cal L}_{\rm m}+{\cal
L}_{\rm int},
\end{equation}
where the electromagnetic part ${\cal L}_{\rm EM}$ has the standard
form
$
{\cal L}_{\rm EM}=\frac{1}{2}\varepsilon_0{\bf E}^2({\bf x},t)-\frac{1}{2\mu_0}{\bf
B}^2({\bf x},t).
$
There is gauge freedom to write the electric field ${\bf E} = -
\partial {\bf A}/\partial t - \nabla \phi$ and the
magnetic field ${\bf B} = \nabla \times {\bf A} $ in terms of the
scalar and vector potentials $\phi$ and $\bf A$. For convenience we
choose the Weyl gauge in which the scalar potential vanishes, which
allows us to write ${\bf E}$ and ${\bf B}$ in terms of only the
vector potential. The amplifying magnetodielectric medium
is modeled with frequency continua of independent vector fields
${\bf X}_\omega({\bf x},t)$ and ${\bf Y}_\omega({\bf x},t)$, in
terms of which we will describe the linear electric and magnetic
polarization of the medium.
Therefore the material part of the Lagrangian density describing the
amplifying medium can now be written as
\begin{subequations}\label{medium Lgrangian}
\begin{eqnarray}
{\cal L}_{\rm e}&=&\int_0^\infty \mbox{d}\omega
\,\left[{\frac{1}{2}{\dot{\bf X}^2}_\omega({\bf
x},t)-\frac{1}{2}\omega^2{\bf X}^2_\omega({\bf
x},t)}\right]\,{\rm sgn}[ \varepsilon_{\rm I}(\omega)],\nonumber\\
\\%
{\cal L}_{\rm m}&=&\int_0^\infty
\mbox{d}\omega\,\left[{\frac{1}{2}{\dot{\bf Y}^2}_\omega({\bf
x},t)-\frac{1}{2}\omega^2{\bf Y}^2_\omega({\bf x},t)}\right]\,{\rm
sgn}[\mu_{\rm I}(\omega)].\nonumber\\
\end{eqnarray}
\end{subequations}
In these Lagrangian densities, for frequencies with positive
signs the medium is lossy, and otherwise it is amplifying as
modeled with inverted oscillators due to the minus sign.
As we will see in Sec.~\ref{Sec:canonical}, this modification leads to field
operators that satisfy Maxwell's equations and whose positive-frequency components are associated with
both annihilation and creation operators in the case of amplifying
media, in accordance with the
previous works~\cite{Matloob1997a,Scheel1998a,Raabe2008a} .

We define the polarization and magnetization fields of the medium as
\begin{subequations}\label{definition of polarization}
\begin{eqnarray}
{\bf P}({\bf x},t)=\int_0^\infty \mbox{d}\omega \, f({\bf x},\omega){\bf
X}_\omega({\bf x},t),\label{4/1}\\
{\bf M}({\bf x},t)=\int_0^\infty \mbox{d}\omega \, g({\bf x},\omega){\bf
Y}_\omega({\bf x},t), \label{4/2}
\end{eqnarray}
\end{subequations}
and assume a linear coupling of the electromagnetic field with these
fields,
\begin{equation}\label{interaction Lagrangian}
{\cal L}_{\rm int}({\bf A},{\bf P},{\bf M}) = {\bf A}({\bf
x},t)\cdot \dot{\bf P}({\bf x},t)+ \nabla\times{\bf A}({\bf
x},t)\cdot{\bf M}({\bf x},t).
\end{equation}
The $f({\bf x},\omega) $ and $g({\bf x},\omega) $ in
Eq.~(\ref{definition of polarization}) are the real-valued scalar
coupling functions of the inhomogeneous medium and the
electromagnetic field.
We have implicity assumed that the medium is isotropic by taking
scalar coupling functions. Anisotropy could be included by making
them tensors, but this is not pursued here.

With the Lagrangian in a suitable form, we can now define a
generating function for our path-integral quantization. For a field
theory with only a single scalar canonical field $\varphi$, the
generating functional (or partition function) has the
form~\cite{Ryder1996a}
\begin{equation}\label{generating functional}
Z[J] = \int {\cal {D}} [\varphi] \exp{\left\{ {\frac{\imath}{\hbar}\int \mbox{d}^4
x[{\cal L}(\varphi (x)) + J(x)\varphi (x)]} \right\}},
\end{equation}
where $x\in \mathbb{R} ^{4}$ is a spacetime coordinate, and $J$ is
the auxiliary source field associated with the scalar field
$\varphi$.
In our case we have several interacting canonical fields in our total
Lagrangian~(\ref{total Lagrangian}), so we need to generalize Eq.~(\ref{generating
functional}). We first calculate the partition function $Z_{0}$ for
the free fields, {\em i.e.} neglecting their interactions:
\begin{eqnarray}\label{free generating functional}
Z_0 [\textbf{J}_{\rm EM},\textbf{J}_{{\rm
e},\omega},\textbf{J}_{{\rm m },\omega}] = \int {\cal{D}}[{\bf A}]
{\cal{D}}[{\bf X}_{\omega}]{\cal{D}}[{\bf Y}_{\omega}]
\nonumber\\
\times\exp\left\{{ \frac{\imath}{\hbar} \int \mbox{d}^4 x \left[{ {\cal{L}}_{\rm EM} +
{\cal{L}}_{\rm  e}+ {\cal{L}}_{\rm m} + \textbf{J}_0\cdot {\bf
A} } \right.} \right.\nonumber\\
\left. { \left. {+\int \mbox{d}\omega \,\textbf{J}_{{\rm e},\omega}\cdot {\bf
X}_{\omega}+\textbf{J}_{{\rm m},\omega}\cdot {\bf
Y}_{\omega}}\right]}\right\},
\end{eqnarray}
in terms of the auxiliary source vector field $\textbf{J}_{\rm EM}$
for the electromagnetic field, and the frequency continua of source
fields $\textbf{J}_{{\rm e},\omega}$ and $\textbf{J}_{{\rm m
},\omega}$ associated with the  electric and magnetic polarization
fields, respectively. We can write $Z_{0}$ in a more convenient form
by employing the four-dimensional version of Gauss's theorem for the
vector potential, thereby replacing ${\cal L}_{\rm EM}$ in Eq.~(\ref{free generating functional}) by
$\left[-\mu_0^{-1}{\bf A}\cdot(\nabla\times\nabla\times{\bf A})-\varepsilon_0{\bf
A}\cdot\partial ^2_t{\bf A}\right]$,
and by using integration by parts for the polarization fields $\bf
X_\omega$ and $\bf Y_\omega$. This gives
\begin{widetext}
\begin{eqnarray}
Z_0 [\textbf{J}_{\rm EM} ,\textbf{J}_{{\rm e},\omega}
,\textbf{J}_{{\rm m},\omega} ] &=& \int {\cal{D}}[{\bf A}]
{\cal{D}}[{\bf X}_\omega]{\cal{D}}[{\bf Y}_\omega] \exp
\left\{{-\frac{\imath}{\hbar} \int \mbox{d}^4 x \,\left[{ \frac{1}{2}{\bf
A}\cdot({\mu_0}^{-1}{\nabla\times\nabla\times}+\varepsilon_0\partial
^2_t){\bf A} -\textbf{J}_{\rm EM} (x)\cdot{\bf A} (x)}\right.}\right.\nonumber\\
&-&\int \mbox{d}\omega \left( {\textbf{J}_{{\rm e},\omega}({ x})\cdot {\bf
X}_{\omega}(x)+\textbf{J}_{{\rm m},\omega}({ x})\cdot {\bf
Y}_{\omega}(x)+ \frac{1}{2}{\bf X}_{\omega}(x)\cdot(\partial
^2_t - \omega^2){\bf X}_{\omega}(x){\rm
sgn}[\varepsilon_{\rm
I}(\omega)]}\right.\nonumber\\
&+&\left.{\left.{\left.{\frac{1}{2}{\bf
Y}_{\omega}(x)\cdot(\partial
^2_t -\omega^2){\bf Y}_{\omega}(x){\rm sgn}[\mu_{\rm
I}(\omega)]}\right)}\right]}\right\}.\nonumber
\end{eqnarray}
\end{widetext}
The above partition function is Gaussian since the integrand is
quadratic in terms of the fields. Therefore the functional
integration can be performed exactly and the result is
$ Z_0 [\textbf{J}_{\rm EM} ,\textbf{J}_{{\rm e},\omega}
,\textbf{J}_{{\rm m},\omega}]
=\exp \bigl\{{\rm i}{\cal L}_{0}/2\hbar\bigl\}$,
in terms of the Lagrangian for non-interacting fields
\begin{widetext}
\begin{eqnarray}
{\cal L}_0&=&\int \mbox{d}^{4}x \int \mbox{d}^4 x'\,\textbf{J}_{\rm
EM}(x) \cdot \bfsfG_{\rm EM}^{(0)} (x - x')\cdot \textbf{J}_{\rm
EM}(x') +\int \mbox{d}^3{\bf x}
\int \mbox{d}t  \int \mbox{d}t' \int \mbox{d}\omega\,\nonumber\\
&&\left\{{\textbf{J}_{{\rm e},\omega}({\bf x},t)\cdot \bfsfG_{{\rm
e},\omega} (t - t')\cdot\textbf{J}_{{\rm e},\omega}({\bf x},t'){\rm
sgn}[\varepsilon_{\rm I}(\omega)]
+\textbf{J}_{{\rm m},\omega}({\bf
x},t)\cdot \bfsfG_{{\rm m},\omega}(t - t')\cdot\textbf{J}_{{\rm
m},\omega}({\bf x},t'){\rm sgn}[\mu_{\rm I}(\omega)] }\right\}.
\end{eqnarray}
\end{widetext}
Here, the space component of the space-time $x\in \mathbb{R}^{4}$ is
indicated in bold by ${\bf {x}}\in \mathbb{R}^{3}$ and the time
component by $t \in \mathbb{R}$. The  Green tensor
 $\bfsfG_{\rm EM}^{(0)} (x- x')$ for the free electromagnetic field satisfies
\begin{equation}
\left(\,\nabla\times\nabla\times+\,\,\frac{1}{c^2} \frac{{\partial ^2
}}{{\partial t^2 }}\,\right) \bfsfG_{\rm EM}^{(0)}(x - x') =\mu_0
\delta^{4}(x - x'),
\end{equation}
describing the propagation of light in time and in free space, whereas
the Green tensors  $\bfsfG_{{\rm e},\omega} (t - t')=\bfsfG_{{\rm m},\omega} (t - t')$ for the
non-interacting electric and magnetic polarization fields describe propagation only in time,
\begin{equation}\label{differential equation for medium field}
\left(\frac{{\partial ^2 }}{{\partial t^2 }} +  \omega ^2 \right)
\bfsfG_{{\rm e/m},\omega} (t - t') = \delta (t - t'){\bm 1}_{3},
\end{equation}
where  ${\bm 1}_{3}$ denotes the spatial unit tensor. This indicates
that the only way to transport energy in the interacting system is
via the electromagnetic field. For the same reason, the polarization
and magnetization fields in the absence of the electromagnetic field
do not lead to a Casimir force~\cite{Kheirandish2010a}. The retarded
solution of Eq.~(\ref{differential equation for medium field}) in
Fourier space  is $\bfsfG_{{\rm e/m},\omega}(\omega')={\bm
1}_{3}/[\omega^2-(\omega'+\imath 0^+)^2]$.
It is convenient to define the source fields ${\bf J}_{\rm P,M}$ for
the electric and magnetic polarization fields as linear combinations
of the corresponding frequency continua ${\bf J}_{\rm e,m}$, namely
${\bf J}_{\rm P}({\bf x}) \equiv \int_0^\infty \mbox{d}\omega
\, f({\bf x},\omega)
{\bf J}_{{\rm e},\omega}({\bf x},\omega)
$ and
${\bf J}_{\rm M}({\bf x}) \equiv \int_0^\infty \mbox{d}\omega
\, g({\bf x},\omega) {\bf J}_{{\rm m},\omega}({\bf x},\omega)
$.
With these, the generating functional for the interacting fields can
be written in terms of the free generating functional as
\cite{Ryder1996a}
\begin{eqnarray}
&&Z[{\bf J}_{\rm EM},{\bf J}_{\rm P},{\bf J}_{\rm M}]=
Z^{-1}_{0}[{\bm 0},{\bm 0},{\bm 0}]\nonumber\\
&&\times \exp \left\{{\frac{\imath}{\hbar}\int
\mbox{d}^4 x\,{\cal{L}}_{\rm int} \left(\frac{\hbar}{\imath}\frac{\delta }
{\delta{\bf J}_{\rm EM} (x)},\frac{\hbar}{\imath}\frac{\delta }{\delta
{\bf J}_{\rm P}(x) },\frac{\hbar}{\imath}\frac{\delta }{\delta
{\bf J}_{\rm M}(x)  }\right)}\right\} \nonumber\\
&& \times Z_{0}[{\bf J}_{\rm EM} ,{\bf J}_{\rm P},{\bf J}_{\rm M} ]
\end{eqnarray}
where $Z_{0}[{\bf J}_{\rm EM} ,{\bf J}_{\rm P},{\bf J}_{\rm M} ]$ is
the free-space partition function and ${\cal{L}}_{\rm int}$ is given
in Eq.~(\ref{interaction Lagrangian}) and $Z^{ - 1}_{0} [{\bm
0},{\bm 0},{\bm 0}]$ is normalization factor. The exponential in
this functional is to be understood as a power series in the
coupling functions, that is by perturbation theory. By using the
specific form Eq.~(\ref{interaction Lagrangian}) for the
interaction, we obtain
\begin{eqnarray}\label{expansion of generating functional for the interacting fields}
&&Z[{\bf J}_{\rm EM},{\bf J}_{\rm P},{\bf J}_{\rm M}] =
Z^{-1}_{0}[{\bm 0},{\bm 0},{\bm 0}] \sum\limits_{n = 0}^\infty
\frac{1}{n!}\biggl\{-\imath \hbar\int \mbox{d}^4x \nonumber\\
&& \left[ {\frac{\delta }
{{\delta {\bf J}_{\rm EM}(x) }} \cdot \frac{\partial }{{\partial
t}}\frac{\delta }{{\delta {\bf
J}_{\rm P}(x) }}+ \,\nabla\times\frac{\delta } {{\delta {\bf J}_{\rm EM}(x)
}} \cdot \frac{\delta }{{\delta {\bf J}_{\rm M}(x)  } }} \right]
\,\,\biggl\}^{n}\nonumber\\
&&\times Z_0 [{\bf J}_{\rm EM} ,{\bf J}_{\rm P},{\bf
J}_{\rm M}].
\end{eqnarray}
Hereby we determined as one of our main results the partition function $Z$ for the interacting
fields that describe a magnetodielectric medium with both linear gain and loss.

\section{Green tensors and susceptibilities of the amplifying medium
}\label{Sec:Greentensors} As we will see below, the Casimir force
can be computed in terms of the electromagnetic Green tensor of the
medium. In our zero-temperature field theory, the Green tensors (or
propagators) are  vacuum expectation values of time-ordered products
of field operators, which can be computed as functional derivatives
of the partition function (see also~\cite{Kheirandish2010a})
\begin{eqnarray}\label{definition of Green function}
\hspace{-.3cm}\bfsfG_{{\rm EM}}(x,x') = -{\rm i}\hbar\frac{{\delta ^2 Z[{\bf J}_{\rm
EM},{\bf J}_{\rm P},{\bf J}_{\rm M}]}}{{\delta {\bf J}_{\rm EM}
(x)\delta {\bf J}_{\rm EM} (x')}}\big |_{{\bf J}_{\rm EM} = {\bf
J}_{\rm P}={\bf J}_{\rm M}={\bm 0}},
\end{eqnarray}
in terms of spacetime coordinates $x,x'$. The medium described by
our Lagrangian~(\ref{total Lagrangian}) in general is not
translationally invariant, but it is stationary, and consequently
$\bfsfG_{{\rm EM}}(x,x')= \bfsfG_{{\rm EM}}({\bf x},{\bf x}',
t-t')$. After evaluating the functional derivatives of
Eq.~(\ref{definition of Green function}) we obtain a Dyson equation
for the Green tensor that after time-Fourier transformation becomes
\begin{widetext}
\begin{eqnarray}\label{Green function for interaction field}
\bfsfG_{\rm EM}({\bf x},{\bf x}',\omega)&=&
\bfsfG_{\rm EM}^{(0)} ({\bf x} - {\bf x}',\omega)\nonumber\\
&+&\omega^2 \int \mbox{d}{\bf x}_1 [\bfsfG_{\rm EM}^{(0)} ({\bf x} -
{\bf x}_1,\omega )\cdot \int \mbox{d}\omega'\{{\rm sgn}(\varepsilon_{\rm
I}(\omega'))f^{2}({\bf x}_1,\omega') \bfsfG_{{\rm e},\omega'} (\omega ) \}\cdot\bfsfG_{\rm EM}
({\bf x}_1,{\bf x}',\omega)]\nonumber\\
&+& \int \mbox{d}{\bf x}_1 [ \bfsfG_{\rm EM}^{(0)} ({\bf x} - {\bf
x}_1,\omega )\times\overleftarrow{\nabla}_1 \cdot\int
\mbox{d}\omega'\{{\rm sgn}(\mu_{\rm I}(\omega')) g^{2}({\bf x}_1,\omega' )  \bfsfG_{{\rm m},\omega'}
(\omega ) \}\cdot \nabla_1\times
\bfsfG_{\rm EM} ({\bf
x}_1, {\bf x}',\omega)].
\end{eqnarray}
\end{widetext}
This long equation for the Green tensor can be brought into a more
familiar form by applying the differential operator
$(\mu_0^{-1}\nabla\times\nabla\times-\omega^2 \varepsilon_0 {\bm 1}_{3})$ to both sides,
giving
\begin{eqnarray}\label{differential equation for Green function}
&&\nabla \times \big[  \mu^{-1}({\bf x},\omega) \nabla  \times
\bfsfG_{\rm EM} ({\bf x},{\bf x}',\omega)\big] \nonumber\\
&&- \frac{\omega ^2
\varepsilon({\bf x},\omega)}{c^2}\bfsfG_{\rm EM} ({\bf x},{\bf x}',\omega)= \mu_0 \delta^{3}({\bf x}-{\bf x}'){\bm 1}_{3}.
\end{eqnarray}
Here we defined the electric permittivity $\varepsilon({\bf
x},\omega)= 1+\chi_{\rm e}({\bf x},\omega)$ and the inverse magnetic
permeability $\mu^{-1}({\bf x},\omega)=1-\chi_{\rm m}({\bf
x},\omega)$  of the amplifying magnetodielectric via
\begin{subequations}\label{susceptibility function in frequency domain}
\begin{eqnarray}
&&\chi_{\rm e}({\bf x},\omega) = \frac{1}{\varepsilon_0}\int_0^\infty \mbox{d}\omega'
\frac{f^{2}({\bf x},\omega' ){\rm sgn}[\varepsilon_{\rm
I}(\omega')]}{\omega'^2-(\omega+\imath0^+)^2},\label{a26/1}\\
&&\chi_{\rm m}({\bf x},\omega) = \mu_0 \int_0^\infty
\mbox{d}\omega'\frac{g^{2}({\bf x},\omega' ) {\rm sgn}[\mu_{\rm
I}(\omega')]}{\omega'^2-(\omega+\imath0^+)^2}.\label{a26/2}
\end{eqnarray}
\end{subequations}
With these definitions and the requirement on the coupling functions that $f^2({\bf x},-\omega^*)=f^2({\bf x},\omega)$ and
$g^2({\bf x},-\omega^*)=g^2({\bf x},\omega)$, the $\varepsilon$ and $\mu$  are complex
functions of frequency which satisfy Kramers--Kronig
relations~\cite{Burnham1969a} and have the properties of the
response functions i.e,
$\varepsilon({\bf x},-\omega^*)=\varepsilon^*({\bf x},\omega)$, and analogously for
$\mu$. Purely lossy media would have the
further properties $\varepsilon_{\rm I}({\bf x},\omega)>0$ and
$\mu_{\rm I}({\bf x},\omega)>0$, but here we have a model that can describe
amplification in some frequency interval(s) as well, for which $
\varepsilon_{\rm I}({\bf x},\omega)<0$ and/or $\mu_{\rm I}({\bf x},\omega)<0$. The functions $\chi_{\rm e,m}({\bf x},\omega)$ have no poles in the upper-half frequency plane and
tend to zero as $\omega\rightarrow\infty$, so that
in the time domain, the electric and magnetic
susceptibilities $\chi_{\rm e,m}({\bf x},t)$ corresponding to
Eq.~(\ref{susceptibility function in frequency domain}) become
proportional to the step function $\Theta(t)$.
This is as it should be,
since either with gain or loss, the response should be causal.

It is important to stress the generality of our model: if we are
given definite functions for the electric permittivity
$\varepsilon({\bf x},\omega)$ and magnetic permeability $\mu({\bf
x},\omega)$ of the gain medium, then we can invert the relations
(\ref{susceptibility function in frequency domain}) to find the
corresponding coupling functions
$f({\bf x},\omega) = \sqrt{2\omega \varepsilon_0 |\varepsilon_{\rm
I}({\bf x},\omega)|/\pi}$
and
$g({\bf x},\omega) = \sqrt{2\omega |\mu^{-1}_{\rm
I}({\bf x},\omega)|/\pi\mu_0 }$,
where the modulus signs ensure that the coupling functions are
real-valued both for lossy and for amplifying media. A similar
general theory, albeit for purely lossy media, can be found
in Refs.~\cite{Philbin2010a,Kheirandish2010b}. Specific choices for the
optical functions $\varepsilon$ and $\mu$ will be made for our
numerical investigations in Sec.~\ref{Sec:Numerical} below.

We will make some further consistency checks on the path-integral quantization for amplifying dielectrics. Similar checks for lossy media were performed in Ref.~\cite{Kheirandish2010a}.
Recall that the defining equation~(\ref{differential equation for Green
function}) for electromagnetic Green tensor is found by functional differentiation of the partition function. This equation enables the
identification of the dielectric functions $\varepsilon(\omega)$
and $\mu(\omega)$ for the amplifying medium. Analogously, we can find Green tensors for
the material fields in our theory, as well as correlations functions of
mixed type. An example of the latter type is
\begin{eqnarray}\label{Green function for polarization}
\bfsfG_{\rm EM,P}({\bf x},{\bf x}',\omega)&=&-{\rm i}\hbar\frac{{\delta
^2 Z[{\bf J}_{\rm EM},{\bf J}_{\rm P},{\bf J}_{\rm M}]}}{{\delta
{\bf J}_{\rm EM} (x)\delta {\bf J}_{\rm P} (x')}}\big
|_{{\bf J}_{\rm EM},\,{\bf J}_{\rm P},{\bf J}_{\rm M}={\bm 0}}\nonumber\\
&=&{\rm i}\omega\varepsilon_0\left[\varepsilon({\bf x},\omega)-1\right]
\bfsfG_{\rm EM}({\bf x},{\bf x}',\omega).
\end{eqnarray}
Analogously we find $\bfsfG_{\rm EM,M}({\bf x},{\bf
x}',\omega)=\mu_0^{-1}\left[1-\mu^{-1}({\bf x},\omega)\right]\nabla\times\bfsfG_{\rm
EM}({\bf x},{\bf x}',\omega)$, with $\varepsilon(\omega)$ and $\mu(\omega)$
as previously defined in Eqs.~(\ref{susceptibility function in frequency domain}).
This shows that the definition of these response functions for amplifying media can be made uniquely and consistently in the path-integral quantization method.
In Sec.~\ref{Sec:canonical} we will derive the equivalent canonical quantization theory for amplifying dielectrics, with the same Lagrangian~(\ref{total Lagrangian}) as a starting point.


\section{Casimir force for amplifying multilayer media}\label{Sec:Casimir}
\subsection{Derivation of $F_{\rm C}$ in path-integral formalism}\label{Sec:Casimir_general}
Here we calculate the Casimir
force for two parallel perfectly conducting
plates that are separated by a multilayer linearly amplifying medium of total width $d$.  Of course perfect conductors do not exist, and the assumption of linear amplification in reality will break down in an amplifying medium without round-trip losses.
Still, an important advantage of our model is that we consider causal optical response functions $\varepsilon(\omega)$ and $\mu(\omega)$, satisfying the Kramers--Kronig relations, as it should for any medium that respects causality, amplifying or not.

For the single homogeneous amplifying medium, the Casimir force between two plates can be computed as the spatial
derivative of the effective action
\begin{equation}\label{F_C_action}
F_{\rm C} = \frac{{\partial }}{{\partial d}} S_{\rm eff} (d),
\end{equation}
where the effective action is proportional to the logarithm of the partition function,
\begin{equation}\label{effective action}
S_{\rm eff} (d) = \hbar\ln Z[d].
\end{equation}
Now since for planar structures there are independent TE and TM
solutions of Maxwell's equations, the total partition function is
the product of $Z_{\rm TE}[d]$ and $Z_{\rm TM}[d]$ partition
functions, so that the effective action intuitively becomes the sum
of TE and TM contributions
\begin{equation}\label{total effective action}
S_{\rm eff}[d]  =\hbar( \ln Z_{\rm TM} [d]+ \ln Z_{\rm TE} [d]),
\end{equation}
and likewise for the Casimir force. The details of the calculation of the
partition functions are left to Appendix~\ref{Sec:Boundaries}, and
yield
\begin{equation}\label{ZTETM}
Z_{\rm TE,TM}  = \frac{1}{{\sqrt {\det \Gamma_{\rm TE,TM} (x,y,z_1,z_2)}
}},
\end{equation}
where
\begin{widetext}
\begin{subequations}\label{GammaTETM}
\begin{eqnarray}
\Gamma_{\rm TM} (x,y,z_1,z_2) &=&\left[{ \begin{array}{*{20}c}
{{\cal G}_{\rm TM}(x - y,z_1,z_1)} & {{\cal G}_{\rm TM}(x - y,z_2,z_1)}  \\
{{\cal G}_{\rm TM}(x - y,z_1,z_2)} & {{\cal G}_{\rm TM}(x - y,z_2,z_2)}  \\
\end{array}}\right],\label{46}\\
 \Gamma_{\rm TE} (x,y,z_1,z_2) &=&\left[{\begin{array}{*{20}c}
-\partial^2_z{{\cal G}_{\rm TE}(x - y,z_1,z_1)} & -\partial^2_z{{\cal G}_{\rm TE}(x - y,z_2,z_1)}  \\
-\partial^2_z{{\cal G}_{\rm TE}(x - y,z_1,z_2)} & -\partial^2_z{{\cal G}_{\rm TE}(x - y,z_2,z_2)}  \\
\end{array}}\right].\label{47}
\end{eqnarray}
\end{subequations}
\end{widetext}
We see from Eqs.~(\ref{F_C_action}-\ref{GammaTETM}) how the Casimir force is expressed in terms of Green tensors ${\cal G}_{\rm TE,TM}$, which can be obtained from the
 Green tensor~Eq.~(\ref{differential equation for Green function}) by applying a Wick rotation. The explicit form
of the Green tensor for planar multilayer dielectric structures were obtained in Ref.~\cite{Tomas1995a}. The details of the calculation of the Green
tensors for the more general situation of amplifying magnetodielectric multilayer media are summarized in Appendix~\ref{Sec:Multilayer}.

Before considering multilayer media in more detail below, we focus on Casimir force in the presence of a single homogeneous amplifying layer.
This is the geometry originally studied by Casimir, but now with the vacuum between the conductors replaced by the amplifying medium.
For this simple geometry, the Dirichlet and Neumann boundary conditions are formally the
same and lead to the same result, so that TE and TM waves each account for half of the Casimir force.

For this case, the Green function~(\ref{differential equation for Green function}) in 2D Fourier space and after Wick rotation
can be written as
${{\cal G}_{\rm EM}({\bf q},{\rm i}\omega,z,z')}= \mu_0 \mu({\rm
i}\omega)\, e\,^{ - {\cal Q }|z - z'|} /(2{\cal Q }),
$
where $
{\cal Q}({\bf q},{\rm i}\omega) =  \sqrt {q^2+\omega^2
\varepsilon(\imath \omega )\mu(\imath \omega)/c^2}$
(see Appendix~\ref{Sec:Multilayer}).

There is no ambiguity how this square root is to be taken, since $\varepsilon(\imath \omega )$ and $\mu(\imath \omega)$ are both real-valued functions of (real) $\omega$. Since both $\varepsilon(\omega )$ and $\mu(\omega)$ have no odd-order zeroes in the upper-half frequency plane and tend to unity in the limit of $|\omega|$ going to infinity, it follows that both $\varepsilon$ and $\mu$ assume positive real values on the positive imaginary frequency axis~\cite{Landau1980a}. Since consequently $\varepsilon(\omega)\mu(\omega)$ does not have any
poles or odd-order zeros in the upper half-plane of frequency, then $n(\omega)$ for ${\rm Im}(\omega)>0$ is defined as
the analytic branch of $\sqrt{\varepsilon(\omega)\mu(\omega)}$ that tends to $+1$ as
$|\omega|$ goes to infinity. (Otherwise, $\sqrt{\varepsilon(\omega)\mu(\omega)}$ would not be
an analytic function there and corresponds to materials with so-called absolute instabilities~\cite{Skaar2006b,Skaar2006a}.) So we find that both $n(\omega)$ and the Green tensor~${{\cal G}_{\rm EM}({\bf q},\omega,z,z')}$ are analytic in the whole upper complex-frequency plane, as it should~\cite{Akyurtlu2010a}. By substituting the expression for the Green function ${{\cal G}_{\rm EM}({\bf q},{\rm i}\omega,z,z')}$  into Eq.~(\ref{effective
action}), the Casimir force per unit area for a homogeneous amplifying medium becomes
\begin{eqnarray}\label{Casimir force for one slab}
F_{\rm C} &=& - \hbar\int \frac{{\mbox{d}\omega \mbox{d}^2 {\bf q}}}{{(2\pi )^3 }}\frac{ 2{\cal
Q}({\bf q},{\rm i}\omega)}{{e^{2{{\cal Q}({\bf q},{\rm i}\omega)}d}  - 1}}\nonumber\\
&=& -\frac{\hbar}{3 c^3\pi^2}\int_0^\infty {\mbox{d}\omega
\,}\frac{\omega\frac{\mbox{d}}{\mbox{d}\omega}\big[n(\imath\omega) \omega\,\big]^3}{{e^{2n(\imath\omega)\omega
d/c\, } - 1}},
\end{eqnarray}
The last identity in~(\ref{Casimir force for one slab})  follows from a partial integration over
$\omega$ and upon calculating the $\omega$-derivative of the
integral over $q$~\cite{Schaden1998a}. For the empty cavity, the
integrals in Eq.~(\ref{Casimir force for one slab}) can be evaluated exactly, giving the
well-known result $F_{\rm C}=-\hbar c \pi^2/(240 d^4)$.

As one of our main results, we find that the Casimir force in the presence of an amplifying
medium~(\ref{Casimir force for one slab}) has the same form as for a
purely attenuating medium~\cite{Kheirandish2010a}, provided that $\varepsilon(\omega)\mu(\omega)$ for amplifying medium does not have any
poles or odd-order zeros in the upper half plane. Thus the actual value of the force for lossy and amplifying media can only follow from the different forms of $\varepsilon( \omega )$ and
$\mu( \omega)$ in both cases. The key difference was already stated in the Introduction, namely that amplifying media have one or more frequency intervals with
$\varepsilon_{\rm I}(\omega)<0$ or $\mu_{\rm I}(\omega)<0$ or both, for real frequencies $\omega>0$, whereas lossy media always have $\varepsilon_{\rm I}(\omega),\mu_{\rm I}(\omega)\ge 0$ for positive real frequencies.

\subsection{Analytical results: bounds on the Casimir force}\label{Sec:Analytical}
Realistic amplifying media are amplifying in one or more frequency intervals and lossy elsewhere. Since the Casimir force~(\ref{Casimir force for one slab}) is obtained as an integral over all frequencies, it may well be that the lossy part dominates the total Casimir force.
We will study these issues numerically in Sec.~\ref{Sec:Numericalgainloss}.
To understand the effect of amplification on the Casimir force,   we will first make the further assumption that the medium is {\em fully amplifying}, by which we mean that $\varepsilon_{\rm I}(\omega),\mu_{\rm I}(\omega)\le 0$ not only for some but for {\em all} positive frequencies. We make this admittedly unrealistic assumption to single out the effect of linear amplification on the Casimir force. This will give us some insight, and after that in Sec.~\ref{Sec:Numerical} we will relax the assumption of full amplification.

Our aim here is to give a bound for the Casimir force in the presence of a fully amplifying medium, similar to the bounds obtained in Ref.~\cite{Lambrecht1997a} for passive dielectric ({\em i.e.} nonmagnetic) mirrors, where from causality considerations it follows that the Casimir force on dielectric slabs is always attractive, but less so than between two ideal mirrors separated by vacuum.

The force~(\ref{Casimir force for one slab}) depends on the refractive index $n(\imath \omega) = \sqrt{\varepsilon(\imath \omega )\mu(\imath \omega)}$, in terms of the susceptibilities $\varepsilon$ and $\mu$ that both tend to unity for high frequencies. Causality implies the identity~\cite{Landau1980a}
\begin{equation}\label{alphaprop}
\varepsilon(i \omega)-1 = \frac{2}{\pi}\int_{0}^{\infty}\mbox{d}\xi\,\frac{\xi \varepsilon_{\rm I}(\xi)}{\omega^{2}+\xi^{2}},
\end{equation}
and an analogous identity holds for $\mu(\omega)$. It follows and is known that for passive systems, with $\varepsilon_{\rm I}(\omega)$ and $\mu_{\rm I}(\omega)$ always positive on the positive real frequency axis, $\varepsilon(\imath \omega)$ and $\mu(\imath \omega)$ decrease monotonically from a finite value $\varepsilon_{\rm static},\mu_{\rm static}>1$ (or $+\infty$ for the electric response of metals) at $\omega=0$ down to unity for $\omega\rightarrow\infty$~\cite{Landau1980a}.

We will instead apply the identity~(\ref{alphaprop}) to amplifying media, for which it also holds as long as they are described by causal response functions. It follows that for fully amplifying systems, with $\varepsilon_{\rm I}(\omega)<0$ and $\mu_{\rm I}(\omega)<0$  on the whole positive real frequency axis, $\varepsilon(\imath \omega)$ and $\mu(\imath \omega)$ from a finite value $<1$ at $\omega=0$ {\em increase monotonically}  towards unity for $\omega\rightarrow\infty$. We can say even more, using that causal response functions $\varepsilon(\omega)$ and $\mu(\omega)$ have no zeroes in the upper-half frequency plane~\cite{Landau1980a}: for fully amplifying media we find that $\varepsilon(\imath \omega)$ and $\mu(\imath \omega)$ increase monotonically from  finite values $0\le \varepsilon_{\rm static},\mu_{\rm static}\le 1$ towards unity for $\omega\rightarrow\infty$.

How are these results related to the Casimir force? We have just found that for fully amplifying media, $n(i \omega)= \sqrt{\varepsilon(\imath \omega )\mu(\imath \omega)}$ increases monotonically, assuming values between $n_{\rm static}\ge 0$ and 1. Therefore, there is a 1-to-1 mapping
\begin{equation}\label{omegas}
\omega \leftrightarrow s \equiv \omega n(i \omega), \quad \omega,s \in [0,\infty).
\end{equation}
This allows us to rewrite the Casimir force for fully amplifying media of Eq.~(\ref{Casimir force for one slab}) in terms of the new variable $s$ as
\begin{equation}\label{Casimiramps}
F_{\rm C} = -\frac{\hbar}{c^{3} \pi^{2}}\int_{0}^{\infty}\mbox{d}s\,\frac{s^{2}\omega(s)}{e^{2 s d/c}-1}.
\end{equation}
Notice that for free space we have $n(i \omega) = 1$ and hence $\omega(s)=s$, which immediately gives the well-known Casimir force for vacuum,
$F_{\rm C}^{\rm vac} = -\hbar c \pi^{2}/(240 d^{4})$.
But also for the general case~(\ref{Casimiramps}) we can say more: since $0\le n_{\rm static} \le n(i \omega) \le 1$ for fully amplifying systems, we can invert the relation ~(\ref{omegas}) and find $s \le \omega(s) \le s/n_{\rm static}$ for all $s$. Combining this with Eq.~(\ref{Casimiramps})~immediately gives for all separations $d$ the inequalities
\begin{equation}\label{bound}
\frac{F_{\rm C}^{\rm vac}}{n_{\rm static}} \le F_{\rm C}^{\rm full \, amp} \le  F_{\rm C}^{\rm vac} = -\frac{\hbar c \pi^{2}}{240 d^{4}}.
\end{equation}
In other words,  the Casimir force on two ideal conductors separated by a fully amplifying medium of width $d$,  is always more attractive than if the medium were vacuum. But it is not more attractive than by a factor $1/n_{\rm static}$.  In particular, we find no sign change in the Casimir force (no Casimir repulsion) on ideal conductors separated by a homogeneous fully amplifying magnetodielectric  medium. These bounds also hold for fully amplifying media that for some frequencies exhibit negative refraction, as numerical examples in Sec.~\ref{Sec:Numerical} will illustrate.

The bound~(\ref{bound}) holds more generally for magnetodielectric media for which $n(\imath \omega)$ increases and $0\le n(\imath \omega)\le 1$. For example, if $\mu(\omega)$ is purely lossy and   $\varepsilon(\omega)$ describes full amplification, then the product of the monotonically decreasing $\mu(\imath \omega)$ and the monotonically increasing $\varepsilon(\imath \omega)$ may still be a monotonically increasing function between 0 and 1.

Similarly, for passive media the causal response functions $\varepsilon(\omega)$ and $\mu(\omega)$ have no zeroes in the upper-half frequency plane and $\varepsilon(\imath \omega)$ and $\mu(\imath \omega)$ decrease monotonically towards unity for $\omega\rightarrow\infty$~\cite{Landau1980a}. Consequently $s/n_{\rm static}\le \omega(s)\le s$ for all $s$, and we find the following inequalities for the Casimir force
\begin{equation}\label{bound_passive}
F_{\rm C}^{\rm vac} \le {F_{\rm C}}^{\rm passive}\le  F_{\rm C}^{\rm vac}/n_{\rm static} \le 0.
\end{equation}
So the Casimir force on two ideal conductors separated by a lossy medium is always attractive, and less attractive than in vacuum, but the force is not reduced by a factor larger than $1/n_{\rm static}$. These bounds also hold for passive media that for some frequencies exhibit negative refraction, as numerical examples in Sec.~\ref{Sec:Numerical} will illustrate.

\subsection{Casimir forces in amplifying multilayer magnetodielectric media}\label{Sec:3layers}
\begin{figure}[t]
\epsfxsize=4cm \ \centerline{\hspace{0cm}\epsfbox{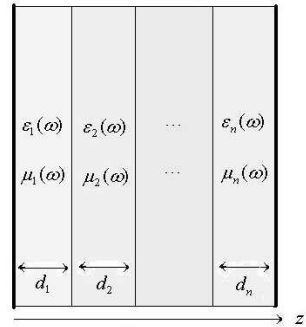}} \
\caption{A planar multilayer magnetodielectric medium
sandwiched between two perfect mirrors, with $N$ parallel planar layers
labeled by $l=1, 2,\ldots, N$. The coordinate system is
chosen such that the layers are perpendicular to the $z$-axis. Layer $j$ has thickness $d_{j}$. Each layer is
assumed to be homogeneous, isotropic and of infinite transverse size
such that $\varepsilon({\bf x},\omega)=\varepsilon_j(\omega)$ and
$\mu({\bf x},\omega)=\mu_j(\omega)$ for ${\bf x}$ in layer $j$. The conductors have coordinates $z_{1}$ and $z_{2}$, and their separation $d$ equals $\sum_{j}^{N}d_{j}$.
}\label{Fig:multilayer}
\end{figure}

Here we generalize our previous results for homogeneous media to planar multilayer geometries, with $N$ parallel planar layers
labeled by $l=1, 2,\ldots, N$ of thicknesses $d_l$,  as depicted in  Fig.~\ref{Fig:multilayer}. Each layer is
assumed to be homogeneous, isotropic and of infinite transverse size.
As is well known for such a planar multilayer geometry, the electromagnetic field
can be completely expanded into independent transverse-electric (TE) and transverse magnetic (TM) fields, that satisfy
the same scalar wave equation, but differ in their boundary conditions. The Green tensor, that via Eq.~(\ref{GammaTETM}) determines the Casimir force, can also be separated into TE and TM parts.
For multilayer dielectric media, the Green tensor was obtained by Toma{\v s}~\cite{Tomas1995a}, essentially using a transfer matrix approach, and a generalization to lossy magnetodielectric  can be found in~\cite{Buhmann2005a}. For our purposes we need a further generalization, namely the Green tensor for  {\em amplifying} magnetodielectric multilayer media, and in Appendix~\ref{Sec:Multilayer} we give a brief derivation and the final result. In general, one finds $N^{2}$ expressions for the Green tensor ${\cal G}^{\rm TE,TM}({\bf q},{\rm i}\omega,z,z')$, depending on which of the $N$ layers the two coordinates $z$ and $z'$ are in, but for the Casimir force on the two ideal conductors, Eq.~(\ref{GammaTETM}) shows that we fortunately only need four of those terms, namely the ones for which $z$ and $z'$ both coincide with one of the coordinates $z_{1}$ (boundary of layer 1) and $z_{2}$ (boundary of layer $N$) of the ideal conductors.

Here we will focus on the Casimir force on two ideal conductors separated by three slabs of matter with linear gain and loss, which are spatially homogeneous in the layers 1, 2, and 3 (see Fig.~\ref{Fig:multilayer}). Following the method outlined in Appendix~\ref{Sec:Multilayer},  whether we find the four relevant expressions for the Green functions, both for ${\rm TE}$ and ${\rm
TM}$ polarizations. We give two of them: For $z,z'$ both in layer 1, we find
\begin{eqnarray}
{\cal G}^{\rm TE,TM}({\bf q},{\rm i}\omega,z,z')&=&\mu_0\mu_1({\rm
i}\omega)\,{\cal I}_{11}^{\rm TE,TM}\nonumber\\
&\times&\frac{e^{-{\cal Q}_1|z-z'|}+r^{\rm TE,TM}_{1+}e^{-{\cal
Q}_1|z+z'|}}{2{\cal Q}_1},\hspace{.6cm}
\end{eqnarray}
while for $z$ in layer 1 and $z'$ in layer 3 we obtain
\begin{eqnarray}
\hspace{-2cm}{\cal G}^{\rm TE,TM}({\bf q},{\rm i}\omega,z,z')&=&\mu_0\mu_1({\rm
i}\omega)\,{\cal I}_{13}^{\rm TE,TM}\nonumber\\
&&\times\frac{t_{1/3}^{\rm TE,TM} e^{{\cal Q}_1z}e^{-{\cal
Q}_3z'}}{2{\cal Q}_1},
\end{eqnarray}
where ${\cal I}_{ij}^{\rm TE}=1$ and ${\cal I}_{ij}^{\rm TM}=\sqrt{\frac{\varepsilon_i\varepsilon_j}{\mu_i\mu_j}}$. Also, the reflection and transmission amplitudes $r^{\rm TE,TM}_{1+}$ and $t_{1/3}^{\rm TE,TM}$ can be calculated
with the recursive relations~(\ref{recursive relations}) and
(\ref{transmision recursive relations}). The other two Green functions can be found analogously, for both polarizations. From
Eq.~(\ref{total effective action}) we then find the effective action for the amplifying three-layer magnetodielectric medium,
\begin{eqnarray}\label{57}
S_{\rm eff}&=&\hbar\int \frac{{\mbox{d}\omega \mbox{d}^2 {\bf q}}}{{(2\pi )^3
}}\left[{\ln \left\{{(1+r_{2,-}^{\rm TM}e^{-2{\cal Q}_1d_1})(1+r_{2,+}^{\rm
TM}e^{-2{\cal Q}_3d_3})}\right.}\right.\nonumber\\
&-&(e^{-2{\cal Q}_1d_1}+r_{2,-}^{\rm
TM})(e^{-2{\cal
Q}_3d_3}+r_{2,+}^{\rm TM})e^{-2{\cal Q}_2d_2}\}\nonumber\\
&+& ({\rm TM}\rightarrow {\rm TE})].
\end{eqnarray}
One can check that in the absence of amplification, our expression~(\ref{57}) tends to a known result for lossy media~\cite{Lischitz1956a,Teo2010a}. In the limit where the two perfect
conductors are brought to infinity (i.e., $d_1,d_{3}\rightarrow
\infty$), Eq.~(\ref{57}) reduces to
\begin{eqnarray}\label{Lifschitz_amplifying}
S_{\rm eff}&=&\hbar\int \frac{{\mbox{d}\omega \mbox{d}^2 {\bf q}}}{{(2\pi )^3 }}\,[\,
\ln\{(1-r_{2,+}^{\rm TM}r_{2,-}^{\rm TM})e^{-2{\cal Q}_2d_2}\}\nonumber\\
&+&({\rm TM}\rightarrow {\rm TE})].
\end{eqnarray}
This is the generalized Lifshitz formula for the Casimir energy
density for amplifying media, in the specific three-layer geometry where two semi-infinite media
with permittivities and permeabilities $\varepsilon_1$, $\mu_1$ and
$\varepsilon_3$, $\mu_3$ are separated by a medium of permittivity
$\varepsilon_2$ and permeability $\mu_2$.

The most realistic special case of Eq.~(\ref{Lifschitz_amplifying}) is the one where the two semi-infinite media are lossy rather than amplifying, and gain occurs for some frequencies in the middle layer.


\section{Numerical results}\label{Sec:Numerical}
\subsection{Numerical results for fully amplifying media}\label{Sec:Numericalfullamp}
\begin{figure}[t]
\epsfxsize=3in \ \centerline{\hspace{0cm}\epsfbox{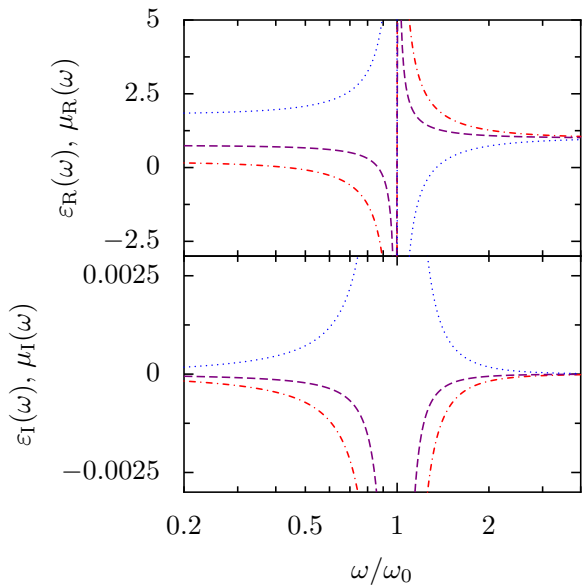}} \
\caption{Frequency dispersion of the real (upper panel) and imaginary (lower panel) parts of the electric permittivity and the magnetic permeability, with (real) frequencies on a log scale. The media are assumed to be single-resonance media with resonance frequency $\omega_0$ and with $\varepsilon(\omega)=\mu(\omega)$ of the form described by Eq.~(\ref{gainmodelLorentz}). Parameters are: for the loss material
(dotted curve): $\omega_{{\rm p
(e, m)}}/\omega_0=0.9,\,\gamma_{{\rm (e, m)}}/\omega_0=0.001$, and for
the gain materials the parameters are similar to the loss material
but with $\omega_{{\rm p (e, m)}}/\omega_0=0.9$ (dotdashed curves),
and $\omega_{{\rm p (e, m)}}/\omega_0=0.5$ (dashed curves). Note that there are frequency
regions where refractive index become negative.} \label{Fig:permittivity in real frequency}
\end{figure}

 In our numerical investigations, we first
choose  homogeneous single-resonance Lorentz-oscillator
models~\cite{Jackson1998a} both for the electric permittivity
$\varepsilon(\omega)$ and for the magnetic permeability $\mu(\omega)$. For simplicity we also assume that the electric and magnetic responses are the same,
\begin{equation}\label{gainmodelLorentz}
\varepsilon(\omega) = \mu(\omega)  =  1-\frac{\omega_{\rm p}^2}{\omega_{0}^2-\omega^2-\imath\gamma\omega},
\end{equation}
where $\omega_{\rm p}$ is the coupling
frequency, $\omega_{0}$ the transverse
resonance frequency, and $\gamma$ the
amplification parameter. The minus sign in front of the second
term in Eq.~(\ref{gainmodelLorentz}) accounts for optical gain that arises
from population inversion in the medium, and it differs from the
usual positive sign for passive systems, consisting for example of two-level systems in their ground
states. The model~(\ref{gainmodelLorentz}) is an example of a fully amplifying medium
(introduced in Sec.~\ref{Sec:Analytical}), since $\varepsilon_{\rm I}(\omega),\mu_{\rm I}(\omega)<0$ for all positive frequencies, as depicted in Fig.~\ref{Fig:permittivity in real frequency}.

In Fig.~\ref{Fig:permittivity in real frequency},
the upper panel depicts the real parts $\varepsilon_{\rm R}(\omega),\mu_{\rm R}(\omega)$. The important thing to notice is that for frequencies $(\omega_0^2-\omega_{\rm p}^2)^{1/2}\lesssim\omega <\omega_ {0}$ both $\varepsilon_{\rm R}(\omega)$ and $\mu_{\rm R}(\omega)$ are negative for amplifying media, whereas for passive media they are negative in the region $\omega_ {0}\leq\omega\lesssim (\omega_0^2+\omega_{\rm p}^2)^{1/2}$. So our model with $\varepsilon(\omega)=\mu(\omega)$ describes  negative refraction, and in particular the perfect-lens situation $n(\omega)=-1$ occurs, or at least $\mbox{Re}[n(\omega)]=-1$. The lower panel of Fig.~\ref{Fig:permittivity in real frequency} depicts the different signs of $\varepsilon_{\rm I}(\omega),\mu_{\rm I}(\omega)$ for lossy and for fully amplifying media.
The requirement that $\varepsilon(\omega)$ and $\mu(\omega)$ have no simple zeroes in the upper-half frequency plane does not restrict any parameters of Lorentz-oscillator models describing loss, but for the fully amplifying model~(\ref{gainmodelLorentz}) we must require $\omega_{\rm p}<\omega_{0}$~\cite{Skaar2006b}. This requirement is usually met, since for natural materials the
permeability of the medium typically equals unity and $\gamma_{\rm e}\ll \omega_{\rm p\,e}\ll\omega_{0{\rm e}}$~\cite{Burnham1969a}.

\begin{figure}[t]
\epsfxsize=3.1in \ \centerline{\hspace{0cm}\epsfbox{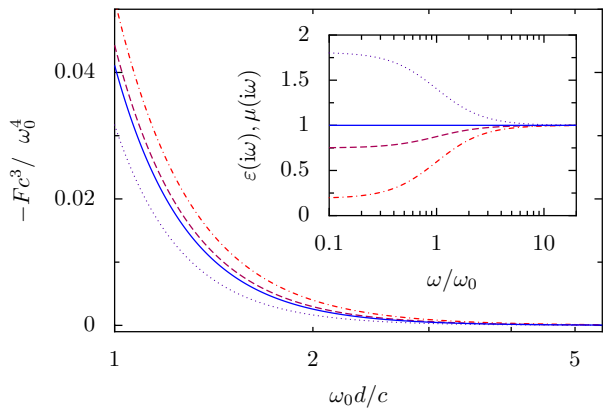}} \
\caption{Casimir force $F_{\rm C}$ per unit area on two perfectly conducting plates, as a function of their separation, for different fully amplifying and lossy dispersive materials
between the plates. The parameters are identical to those used in Fig.~\ref{Fig:permittivity in real frequency}. The solid curve corresponds to vacuum. The force is given in units of  $\hbar\omega_0^4/c^{3}$ and the separation in units of $c/\omega_0$. The inset shows $\varepsilon=\mu$ at imaginary
frequencies. }\label{Fig:Casimir force}
\end{figure}

In Figure~\ref{Fig:Casimir force}
we compare the Casimir force on two perfect planar conductors as described by Eq.~(\ref{Casimir force for one slab}) for
lossy and for fully amplifying homogeneous media, as well as for vacuum, all as a function of plate separation~$d$. The inset~\ref{Fig:Casimir force} shows the real-valued $\varepsilon(\imath \omega)$, which is indeed monotonically decreasing as discussed in Sec.~\ref{Sec:Analytical}, here from $\varepsilon_{\rm static}=1+\omega_{\rm p}^{2}/\omega_{0}^{2}$ down to unity for lossy media. For amplifying media $\varepsilon(\imath \omega)$ is indeed monotonically increasing, from $\varepsilon_{\rm static}=1-\omega_{\rm p}^{2}/\omega_{0}^{2}$ towards unity. For some amplifying negative-index geometries, the Casimir force was found to be repulsive~\cite{Leonhardt2007a}, but Fig.~\ref{Fig:Casimir force} illustrates our finding of Sec.~\ref{Sec:Analytical} for a single homogeneous layer: attractive Casimir forces for all plate separations  for all homogeneous fully amplifying Kramers--Kronig media, including the media where for some frequencies there is negative refraction. More specifically, it is easily verified that the bounds of  Eq.~(\ref{bound}) hold in Fig.~\ref{Fig:Casimir force}: the attraction for fully amplifying media is always stronger than for plates separated by vacuum, but weaker than $1/n_{\rm static}$ times the free-space value. Likewise, the analogous bounds on $F_{\rm C}$ in Eq.~(\ref{bound_passive}) for passive media are also seen to hold in Fig.~\ref{Fig:Casimir force}.

\subsection{Numerical results for homogeneous medium with both gain and loss}
\label{Sec:Numericalgainloss}

\begin{figure}[t]
\epsfxsize=3in \ \centerline{\hspace{0cm}\epsfbox{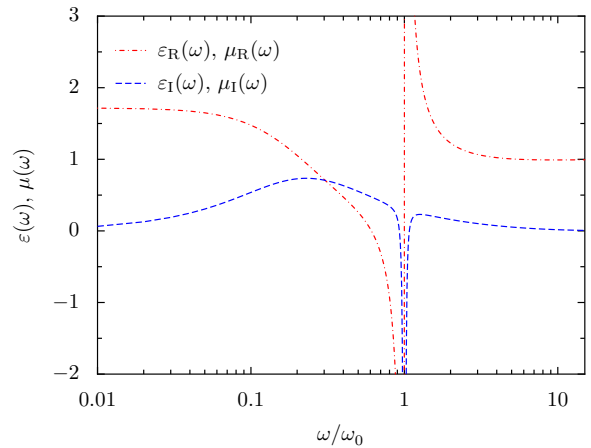}} \
\caption{The real and imaginary parts of the electric permittivity and the magnetic permeability in real frequency for the mixed-type gain material
with electromagnetic parameters described by Eq.~(\ref{epsilon_gain_and_loss})
{\em: $\omega_{0}^{\rm
l}/\omega_{0}^{\rm
g}=1.5$, $\omega_{{\rm p }}^{\rm
g}/\omega_{0}^{\rm
g}=0.85$, $\omega_{{\rm p }}^{\rm
l}/\omega_{0}^{\rm
g}=1.8$, $\gamma^{\rm g}/\omega_{0}^{\rm
g}=0.01$, $\gamma^{\rm l}/\omega_0^{\rm g}=10$}.
}\label{Fig:permittivitygainandloss}
\end{figure}

The fully amplifying media studied in Secs.~\ref{Sec:Analytical} and \ref{Sec:Numericalfullamp} do not occur in nature, but give insight in the effect of amplification on Casimir forces. Real amplifying media are typically amplifying in a limited frequency interval, and lossy elsewhere. Therefore we will now study the effect of gain in a limited frequency interval on the Casimir force, but we still assume $\varepsilon(\omega)=\mu(\omega)$ for simplicity. We modify the single-resonance model~(\ref{gainmodelLorentz}) for fully amplifying media by adding a loss term,
\begin{equation}\label{epsilon_gain_and_loss}
\varepsilon(\omega)
=1-\frac{(\omega_{\rm p}^{{\rm g}})^{2}}{(\omega_{0}^{\rm g})^{2}-\omega^2-\imath\omega\gamma^{{\rm
g}}}
+
\frac{(\omega_{\rm p}^{{\rm l}})^{2}}{(\omega_{0}^{\rm l})^{2}-\omega^2-\imath\omega\gamma^{{\rm l}}}.
\end{equation}
By our choice of parameters, we describe a medium with gain around $\omega_{0}^{\rm g}$ and loss elsewhere, as seen in Fig.~\ref{Fig:permittivitygainandloss}.

\begin{figure}[t]
\epsfxsize=3.5in \ \centerline{\hspace{0cm}\epsfbox{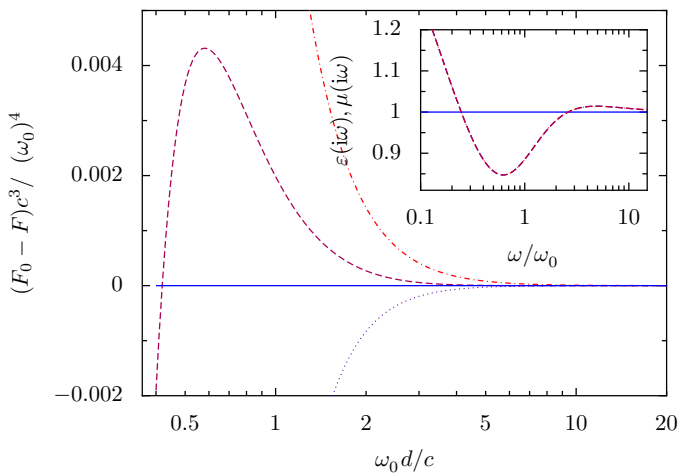}} \
\caption{The difference of the Casimir force with respect to the vacuum value, for several
dispersive media: lossy, fully amplifying (Eq.~(\ref{gainmodelLorentz})), and mixed-type (Eq.~(\ref{epsilon_gain_and_loss})). Scaling of the force and of the distance, and parameters for the fully amplifying
(dot-dashed) and for lossy media (dotted curves) are similar
 as in Fig.~\ref{Fig:permittivity in real frequency}; parameters for mixed-type (dashed curves) as in Fig.~\ref{Fig:permittivitygainandloss}. The solid curve correspond to vacuum. The inset shows the corresponding $\omega$-dispersion of $\varepsilon(\imath \omega )=\mu( \imath\omega)$ for a mixed-type gain medium.
}\label{Fig:gainandloss}
\end{figure}

In Figure~\ref{Fig:gainandloss}, we show the corresponding Casimir force for this medium, as calculated with Eq.~(\ref{Casimir force for one slab}). More precisely, the figure depicts the difference of the Casimir force with respect to the free-space value, and such that a positive value corresponds to a more strongly attractive Casimir force than for free space. For lossy media, we know that this will result in a curve entirely below the horizontal axis, while for fully amplifying media only positive   curves would result, as seen in the figure. For the medium~(\ref{epsilon_gain_and_loss}) with both gain and loss and with $\varepsilon(\omega)=\mu(\omega)$, it follows from the inset of Fig.~\ref{Fig:gainandloss} that  $n(\imath\omega)$ is not monotonically increasing or decreasing, but shows more complex behavior: for low frequencies, the curve decreases as for a purely lossy medium. After going through a minimum, the curve increases for a while as fully amplifying media would do monotonically, and then finally it decreases again.

In the main panel of Fig.~\ref{Fig:gainandloss},
 showing the Casimir force as a function of distance, we see similar behavior: for low frequencies the curve is negative and the force is attractive but weaker than for free space, reminding of a purely lossy medium. For intermediate distances, the force is more attractive than for free space, as we have seen  for fully amplifying media in Secs.~\ref{Sec:Analytical} and~\ref{Sec:Numericalfullamp}. Finally for large distances - but this is not clearly visible in the main graph - $F_{\rm C}$ again becomes weaker than $F_{\rm C}^{\rm vac}$, again reminding of a lossy medium.

 These observations agree with the known fact that at long distances, the main contribution to the Casimir force comes from the low-frequency region, while the force at short distance depends on the high-frequency behavior~\cite{Rosa2008a}. This fact follows from the frequency-integral representation~(\ref{Casimir force for one slab}) of the force. From Fig.~\ref{Fig:Casimir force} for the fully amplifying media this relation was not evident, but in Fig.~\ref{Fig:gainandloss} it is:  gain at a finite frequency interval around $\omega_{0}^{\rm g}$ may lead to Casimir forces  that are more attractive than in free space for a finite interval of plate separations, roughly around a separation $d=\lambda_{0}^{\rm g}/(2\pi)$, in terms of the resonance wavelength $\lambda_{0}^{\rm g} = 2\pi c/\omega_{0}^{\rm g}$. For closer or more distant separations, the lossy character of the gain-and-loss medium dominates the Casimir force, so that $F_{\rm C}^{\rm vac}<F_{\rm C}<0$. As for the fully amplifying media, we do not find Casimir repulsion for homogeneous media with both gain and loss.

\section{Canonical quantization of electromagnetic field in amplifying medium}
\label{Sec:canonical}
In the previous sections we made a shortcut
from the Lagrangian via the partition function and Green tensor to
calculate the Casimir force for amplifying media. To find the Casimir force, we did not need to perform
explicitly a canonical quantization for amplifying media based on
our Lagrangian. The reason to do this here is to make contact with other approaches and to provide an underlying canonical theory for the phenomenological
quantum electrodynamics for amplifying media that was developed in
recent years~\cite{Jeffers1993a,Jeffers1996a,Matloob1997a,Scheel1998a,Raabe2008a}.

The starting point is the Lagrangian~(\ref{total Lagrangian}),
with the vector potential ${\bf A}$, and the continua of
polarization operators ${\bf X}_\omega$ and ${\bf Y}_\omega$ as
canonical fields with the following canonically conjugate fields
\begin{subequations}\label{canonically conjugate fields}
\begin{eqnarray}
- \varepsilon_0 {\bf E}({\bf x},t)& =& \frac{\delta {\cal L}}{\delta \dot{\bf
{A}}({\bf x},t)}= \varepsilon_0{\dot {\bf A}}({\bf x},t),\\
{\bf Q}_\omega({\bf x},t) &=& \frac{\delta {\cal L}}{\delta \dot{\bf
{X}_\omega}({\bf x},t)} \nonumber\\
&=&f(\omega,{\bf x}){\bf A}({\bf x},t)+ {\rm sgn}[\varepsilon_{\rm I}(\omega)]{\dot
{\bf X}}_\omega({\bf x},t), \\
{\bf \Pi}_\omega({\bf x},t) &=& \frac{\delta {\cal L}}{\delta {\bf
\dot{Y}_\omega}({\bf x},t)}\nonumber\\
&=& g(\omega,{\bf x})\nabla\times{\bf A}({\bf x},t)+{\rm sgn}[\mu_{\rm I}(\omega)]{\dot
{\bf Y}}_\omega({\bf x},t).\nonumber\\
\end{eqnarray}
\end{subequations}
So here we find $- \varepsilon_0 {\bf E}$ as the canonical conjugate to the vector potential, as in
Refs.~\cite{Matloob1997a,Scheel1998a,Raabe2008a,Huttner1992a,Wubs2001a,Suttorp2004a}.
Apart from the subtlety with the sign functions in Eqs.~(\ref{canonically conjugate fields}) that discriminate
between the frequency intervals where there is gain and loss, the
canonical quantization of the  fields can proceed in a standard
fashion by demanding equal-time commutation relations among the
variables and their conjugates,
\begin{subequations}\label{equal_time_commutation}
\begin{eqnarray}
\left[ { {A} _i ({\bf x},t),-\varepsilon_0 {E}_j ({\bf x'},t)} \right] &=& {\rm
i} \hbar \,\delta _{ij} \delta ^{\bot}({\bf x} - {\bf x'}),\label{b2}\\
\left[ { {X} _{\omega,i} ({\bf x},t),- {Q}_{\omega',j} ({\bf x'},t)}
\right]& =& {\rm i}\hbar\, {\rm sgn }[\varepsilon_{\rm I}(\omega)]\,\delta
_{ij}
\nonumber\\
&\times&\delta(\omega-\omega') \delta^{3} ({\bf x} - {\bf x'}),\;\;\\
\left[ { {Y} _{\omega,i} ({\bf x},t),- {\Pi}_{\omega',j} ({\bf
x'},t)} \right]& =& {\rm i}\hbar\,{\rm sgn }[\mu_{\rm I}(\omega)]\,\delta _{ij}\nonumber\\
&\times&\delta(\omega-\omega') \delta^{3} ({\bf x} - {\bf
x'}),\label{b4}
\end{eqnarray}
\end{subequations}
with all other equal-time
commutators vanish. Using the Lagrangian~(\ref{total Lagrangian})
and the expression for the canonical conjugate variables in
(\ref{canonically conjugate fields}), we obtain the Hamiltonian
density
\begin{eqnarray} \label{Hamiltonian density}
{\cal H} &=&  \frac{1}{2}\varepsilon_0{\bf E}^2({\bf
x},t)+\frac{{\bf B}^2({\bf x},t)}{2\mu_0}\nonumber\\
&+&\frac{1}{2}\int_0^\infty \mbox{d}\omega\, {\rm
sgn}[\varepsilon_{\rm I}(\omega)]\bigl\{ \dot{\bf {X}}_\omega^2({\bf
x},t)+\omega ^2 {\bf X}_\omega^2({\bf
x},t) \bigl\}\nonumber\\
&+& \frac{1}{2}\int_0^\infty \mbox{d}\omega\, {\rm sgn}[\mu_{\rm
I}(\omega)]\bigl\{ \dot{\bf {Y}}_\omega^2({\bf
x},t)+\omega ^2 {\bf Y}_\omega^2({\bf x},t) \bigl\}.\;\;\;
\end{eqnarray}
Maxwell's equations can now be obtained from the Heisenberg
equations of motion for the vector potential and the transverse electric
field and from the commutation relation
(\ref{equal_time_commutation}),
\begin{subequations}\label{ADderivatives}
\begin{eqnarray}
\dot{\bf {A}}({\bf x},t) & = & - {\bf E}({\bf x},t), \\
\varepsilon_0\dot{\bf {E}}({\bf x},t) & = & \frac{\nabla\times\nabla\times{\bf
A}({\bf x},t)}{\mu_0} -\nabla\times{\bf M}({\bf x},t)-\dot{{\bf P}}({\bf x},t).\nonumber\\
\end{eqnarray}
\end{subequations}
Using the definitions ${\bf D} =  \varepsilon_0{{\bf E}}+{{\bf P}}$ and  ${\bf H} = {\bf B}/\mu_0-{\bf M}$ for the displacement
field and the magnetic field strength, respectively, Eqs.~(\ref{ADderivatives}) result in $\dot{\bf {D}}({\bf
x},t)=\nabla\times{\bf H}({\bf x},t)$ and $\dot{\bf {B}}({\bf
x},t)=-\nabla\times{\bf E}({\bf x},t)$, as expected. In a similar
fashion, the Heisenberg equation of motion for the dynamical
variables ${\bf X}_{\omega}$ and ${\bf Y}_{\omega}$  lead to
\begin{subequations}
\begin{eqnarray}
\ddot{\bf {X}}_\omega({\bf x},t) &  = & -\omega^2{\bf X}_\omega({\bf
x},t) + {\rm sgn}[\varepsilon_{\rm
I}(\omega)] f({\bf x},\omega){\bf E}({\bf x},t),\hspace{.8cm}\\
\ddot{\bf {Y}}_\omega({\bf x},t)  & = &  -\omega^2{\bf
Y}_\omega({\bf x},t) + {\rm sgn}[\mu_{\rm I}(\omega)] g({\bf
x},\omega){\bf B}({\bf x},t),\hspace{.8cm}
\end{eqnarray}
\end{subequations}
with formal solution
\begin{eqnarray}
{\bf X}_\omega ({\bf x},t)  &= & \left({\dot{\bf  X}_\omega ({\bf
x},0)\frac{{\sin \omega t}}{\omega }
+ {\bf X} _\omega ({\bf
x},0)\cos \omega t}\right) \nonumber \\ & + & f({\bf x},\omega){\rm
sgn}[\varepsilon_{\rm I}(\omega)]\int_0^t \mbox{d}t' \frac{{\sin \omega
(t - t')}}{\omega } {\bf {\bf E}}({\bf
x},t'),\hspace{.6cm}
\end{eqnarray}
and likewise for ${\bf Y}_\omega ({\bf x},t)$.
To facilitate the calculations, let us introduce the following
annihilation operators
\begin{subequations}\label{creationandannihilation}
\begin{eqnarray}
d_j ({\bf x},\omega ,t) & = & \frac{1}{ \sqrt {{2\hbar\omega }}}
\left[ {\omega {\rm X}_{\omega,j}({\bf x},t ) + {\rm i}\,{\rm
Q}_{\omega,j}({\bf
x},t)} \right],  \label{b14} \\
b_j  ({\bf x},\omega ,t) &  = & \frac{1}{ \sqrt {{2\hbar\omega }}}
 \left[
{\omega  {\rm Y}_{\omega,j}({\bf x},t ) + {\rm i}\,  {\rm
\Pi}_{\omega,j}({\bf x},t)} \right], \label{b15}
\end{eqnarray}
\end{subequations}
where $j=1,2,3$ labels three orthogonal spatial directions. Their
commutation relations follow immediately from
Eq.~(\ref{equal_time_commutation}),
\begin{subequations}\label{comm_creation_annihilation}
\begin{eqnarray}
\left[ {d_j  ({\bf x},\omega ,t),d_{j'}^ \dag ({\bf x}',\omega ',t)}
\right] & = & {\rm sgn}[ \varepsilon_{\rm I}(\omega)]\,\delta _{jj'}\nonumber\\
&\times&\delta (\omega  - \omega ')\delta ({\bf
x} - {\bf x}'),\; \\
\left[ {b_j  ({\bf x},\omega ,t),b_{j'}^ \dag ({\bf x}',\omega ',t)}
\right] & = & {\rm sgn}[ \mu_{\rm I}(\omega)] \, \delta _{jj'}\nonumber\\
&\times&
\delta (\omega  - \omega ')\delta ({\bf x} - {\bf x}').
\end{eqnarray}
\end{subequations}
Now by inverting the relations~(\ref{creationandannihilation}) and
substituting the result into Eqs.~(\ref{definition of
polarization}), the polarization and magnetization fields of the
magnetodielectric medium can be written in terms of creation and
annihilation operators as
\begin{subequations}\label{PandM}
\begin{eqnarray}
{\bf P}({\bf x},t)&=&\varepsilon_0\int_0^\infty
\mbox{d}t'\,\chi_{\rm e}({\bf x},t-t'){\bf E}({\bf x},t')+{\bf
P}^{\rm N}({\bf x},t), \nonumber \\ \\
{\bf M}({\bf x},t)&=&\frac{1}{\mu_0}\int_0^\infty
\mbox{d}t'\chi_{\rm m}({\bf x},t-t'){\bf B}({\bf x},t')+{\bf
M}^{\rm N}({\bf x},t),\nonumber\\
\end{eqnarray}
\end{subequations}
with susceptibilities $\chi_{\rm m,e}$  as defined in
Eq.~(\ref{susceptibility function in frequency domain}). The fields
${\bf P}^{\rm N}({\bf x},t)$ and ${\bf M}^{\rm N}({\bf x},t)$ are
the electric and magnetic polarization noise densities associated
with absorption and amplification. As in the phenomenological
method, we can separate the noise operators into positive- and
negative-frequency parts ${\bf P}^{\rm N}={\bf P}^{\rm N (+)}+{\bf
P}^{\rm N(-)}$ with ${\bf P}^{\rm N(-)} = [{\bf P}^{\rm N(+)}]^{\dag}$ and analogously for ${\bf M}^{\rm N}$, where
\begin{subequations}\label{Noiseoperator}
\begin{eqnarray}
{\bf P}^{\rm N (+)}({\bf x},t)&=&\int_0^\infty \mbox{d}\omega\,
\sqrt{\frac{\varepsilon_0|\varepsilon_{\rm
I}(\omega)|}{\pi}}\bigl\{d_i({\bf
x},\omega,0)\Theta(\varepsilon_{\rm I}(\omega))\nonumber\\
&&+d_i^\dag({\bf
x},\omega,0)\Theta(-\varepsilon_{\rm I}(\omega))\bigl\}e^{-{\rm i}
\omega t},\\
{\bf M}^{\rm N (+)}({\bf x},t)&=&\int_0^\infty \mbox{d}\omega\,
\sqrt{\frac{|\mu_{\rm I}^{-1}(\omega)|}{\pi\mu_0}}\bigl\{b_i({\bf
x},\omega,0)\Theta(\mu_{\rm I}(\omega))\nonumber\\
&&+ b_i^\dag({\bf
x},\omega,0)\Theta(-\mu_{\rm I}(\omega))\bigl\}e^{-{\rm i} \omega
t}\label{b24},
\end{eqnarray}
\end{subequations}
In fact, the above equations are the starting point in
Refs.~\cite{Matloob1997a,Scheel1998a,Raabe2008a} to the phenomenological quantization of
the electromagnetic field in amplifying magnetodielectric media.

If we now take the time derivative of Eq.~(\ref{ADderivatives}) and use Eq.~(\ref{PandM}), this yields
the frequency-domain wave equation for the positive-frequency
part of the vector potential
\begin{eqnarray}\label{b25}
&&\nabla \times \big[  \mu^{-1}({\bf x},\omega) \nabla  \times {\bf
A}^{ (+)} ({\bf x},\omega)\big] - \frac{\omega ^2}{c^2}
\varepsilon({\bf x},\omega){\bf A}^{ (+)} ({\bf x},\omega) =\nonumber\\
&&-{\rm i}\mu_0\omega{\bf P}^{\rm N (+)}({\bf x})+\mu_0\nabla\times{\bf
M}^{\rm N (+)}({\bf x},\omega).
\end{eqnarray}
This equation can be solved as
\begin{eqnarray}\label{vector potential field}
{\bf A}^{ (+)} ({\bf x},t)=\frac{1}{2\pi}\int_0^\infty \mbox{d}\omega
\int \mbox{d}^3 {\bf x'}\, \bfsfG({\bf x},{\bf x'},\omega) \cdot \nonumber\\
\left[-{\rm i}\omega{\bf P}^{\rm N (+)}({\bf x',\omega})+\nabla\times{\bf
M}^{\rm N (+)}({\bf x'},\omega)\right]\, e^{-{\rm i}\omega t},
\end{eqnarray}
where the Green tensor $\bfsfG({\bf x},{\bf x'},\omega)$ is the
solution of Eq.~(\ref{differential equation for Green function}).

The equations~(\ref{Noiseoperator}) and~(\ref{vector potential field}) and
the commutation relations are the same as
obtained from the phenomenological method~\cite{Jeffers1993a,Jeffers1996a,Matloob1997a,Scheel1998a,Raabe2008a}.
Therefore, with our Lagrangian~(\ref{total Lagrangian}) and the canonical quantization performed here, we formulated a  microscopic
basis for the phenomenological quantization of the electromagnetic field in amplifying magnetodielectric media.


\section{Conclusions and discussion}\label{Sec:Conclusions}
 The electromagnetic field in  an amplifying
magnetodielectric medium was quantized with
a path-integral technique. We determined correlation functions of different fields
and found electric and magnetic
susceptibilities of the amplifying medium that are consistent with causality. We determined the Green functions
in amplifying planar multilayer magnetodielectrics and used this  to calculate the
Casimir energy and force in such media.

The calculations show that the form of the Casimir force as a functional of the dielectric functions $\varepsilon({\bf x},\omega),\,\mu({\bf x},\omega)$ does not change significantly as compared to passive media, but some more caution is needed, especially about the sign of wave-vectors components in amplifying negative-index materials in the direction normal to the planes.

 Here we studied systems that have both gain and a negative index, and to that end we assumed $\varepsilon(\omega)=\mu(\omega)$ in our numerical calculations, not because this would be simple to realize in experiments but rather to gain insight. The concept was introduced of fully amplifying media, {\em i.e.} amplifying at all frequencies. The advantage of our approach is that our optical functions $\varepsilon(\omega),\mu(\omega)$ are defined at all frequencies and are causal. For two conductors separated by a homogeneous passive medium, we find that the Casimir force satisfies the bounds of Eq.~(\ref{bound_passive}), so that it is always attractive and less so than if the medium were replaced by vacuum. On the other hand, for a homogeneous fully amplifying medium, we find the bounds~(\ref{bound}) that the Casimir force is finite, and always {\em more attractive} than in vacuum.

 Both bounds are remarkable insofar that both the fully amplifying and the passive media with $\varepsilon(\omega)=\mu(\omega)$ may have negative refraction in a large frequency interval, whereas for some other planar geometries it was reported that negative refraction may lead to Casimir forces that may become repulsive~\cite{Leonhardt2007a}, and also that gain would lead to repulsive Casimir forces~\cite{Sherkunov2005a,Leonhardt2007a}. This is not a contradiction, however, but rather shows that negative refractive indices or gain do not automatically imply a sign change of the Casimir force, and that the sign of the force strongly depends on the geometry also for amplifying and negative-index materials.

The bounds~(\ref{bound},\ref{bound_passive}) for homogeneous fully amplifying and for passive media have in common that the force is bound by the free-space Casimir force on the one side and by the free-space Casimir force divided by the static refractive index $n_{\rm static}$ on the other. It does not matter whether for microwaves or optical frequencies the medium has a negative refractive index or not, as long as the static refractive index stays the same. In that sense, the Casimir force has little to do with optics, both for passive and for amplifying media. Ref.~\cite{Rosa2008a} also stresses the importance of the low-frequency behavior, and here we found an illustrative example also for amplifying systems.

Some observables have divergent values in models of linear amplification, especially for geometries with round-trip gain. Our simple geometry of perfect conductors separated by a single amplifying medium will even exhibit round-trip gain for all frequencies for which there is amplification. Nevertheless we find that the Casimir force on the conductors is finite, bounded by the inequalities of Eq.~(\ref{bound}). This could be explained by the fact that the Casimir force is a vacuum force, so that there are no photons present that are amplified indefinitely.

Casimir forces in amplifying media are only beginning to be explored, and we only considered linear amplification. This is not always a realistic model, especially for geometries where the linear-amplification model predicts round-trip gain. It is an open challenge to calculate Casimir forces in the presence of media with nonlinear amplification, for example with parametric amplification. A possible route could be to generalize the Lagrangian for passive nonlinear media of Ref.~\cite{Kheirandisch2011a} to gain media, and the result would be a generalization of our Lagrangian of  Eq.~(\ref{total Lagrangian}) to nonlinearly amplifying media.

Finally, we carried out a canonical quantization of the electromagnetic field in an arbitrary linear amplifying and/or passive medium, and showed that the resulting field operators satisfy the macroscopic Maxwell equations for
an arbitrary linearly amplifying and/or passive medium. The resulting theory is valid for all linear, inhomogeneous, amplifying and/or passive magnetodielectric media with dielectric functions that satisfy the Kramers-–Kronig relations. The postulates of the phenomenological theory~\cite{Jeffers1993a,Jeffers1996a,Matloob1997a,Scheel1998a,Raabe2008a}
that serve as its starting point were
here derived by canonical quantization.

\section*{Aknowledgments}
E. A. would like to thank the Technical University of Denmark for hospitality and the University of Isfahan and the Iranian Ministry of Science, Research, and Technology for financial support. M. W. acknowledges financial support by The Danish Research Council for Technology and Production Sciences (FTP grant $\#274 - 07 - 0080$).


\appendix

\section{Generating function in presence of Dirichlet and Neumann boundary conditions}
\label{Sec:Boundaries}
Here we calculate the generating
function for two perfectly conducting plates surrounding an amplifying planar
multilayer system. We will be brief, as most of the calculation is identical to the case of lossy media, see Ref.~\cite{Kheirandish2010a} for example.

We can consider TE and TM polarized waves separately. On the two plates (labeled by $\alpha$) they satisfy Dirichlet or Neumann
boundary conditions, respectively,
\begin{subequations}\label{The Dirichlet or Neumann boundary conditions}
\begin{eqnarray}
\varphi_{\rm TM}|_{S_\alpha} = 0,\,\,\,\, \alpha=1,2\label{28}\\
{\partial_{\rm n}}\varphi_{\rm TE}|_{S_\alpha} = 0,\,\,\,\,
\alpha=1,2 \label{29}
\end{eqnarray}
\end{subequations}
on each surface ${S_\alpha}$, where ${\partial_{\rm n}} $ is the
normal derivative of the surface ${S_\alpha}$ pointing into the
space between the two plates. To obtain the partition function in 4D Euclidean space from
the Lagrangian~(\ref{total Lagrangian}), we
made a Wick rotation so that the signature of space-time changes from
Minkowski to Euclidean.

In the 4D Euclidean space, the plates are parameterized by ${\cal
X}_1({\bf x},z_1,\imath t)$ and ${\cal X}_2({\bf x},z_2,\imath t)$, where
${\bf x}=(x,y)$. The Dirichlet or Neumann
boundary conditions corresponding to the constraints (\ref{The
Dirichlet or Neumann boundary conditions}) can be imposed by
inserting delta functions which can be expressed in terms of
auxiliary fields $\psi _\alpha ({\cal X}_\alpha )$ as
follows~\cite{Emig2001a,Bordag2009a}
\begin{subequations}\label{delta functions}
\begin{eqnarray}
\delta (\varphi ({\cal X}_\alpha  )) &=& \int {\cal {D}}[\psi_\alpha
({\cal X}_\alpha )]e^{\imath\int \mbox{d}{\cal X}_\alpha \psi ({\cal
X}_\alpha  )\varphi ({\cal X}_\alpha )},\label{30}\\
\delta ({\partial_n}\varphi ({\cal X}_\alpha  )) &=& \int {\cal
{D}}[\psi_\alpha ({\cal X}_\alpha )]e^{\imath\int  \mbox{d}{\cal X}_\alpha
\psi ({\cal X}_\alpha ){\partial_n}\varphi ({\cal X}_\alpha
)}.\hspace{.5cm}\label{31}
\end{eqnarray}
\end{subequations}
Using Eqs.~(\ref{delta functions}), the partition function (\ref{generating functional}) in
Euclidean space can be written as
\begin{subequations}\label{ZTMZTE}
\begin{eqnarray}
Z_{\rm TM} &=& Z_0^{ - 1} \int {\cal {D}}[\varphi] \prod\limits_{^{a
= 1} }^2 \delta (\varphi ({\cal X}_\alpha  ))e^{-S_{\rm E, TM}[\varphi
]},\label{32}\\
Z_{\rm TE}& =& Z_0^{ - 1} \int {\cal {D}}[\varphi] \prod\limits_{^{a
= 1} }^2 \delta ({\partial_{\rm n}}\varphi ({\cal X}_\alpha
))e^{-S_{\rm E, TE}[\varphi ]},\;\label{33}
\end{eqnarray}
\end{subequations}
where the Euclidean actions $ S_{\rm E,TM/TE} (\varphi )$ are defined as
\begin{subequations}\label{Euclidean action}
\begin{eqnarray}
S_{\rm E,TM}[\varphi ] &=& \int \mbox{d}^{4} x \bigl\{{{\cal L} (\varphi(x))}\nonumber\\
&+& {\varphi (x)\sum\limits_{\alpha = 1}^2 \int \mbox{d}{\cal X}\delta ({\cal X} - {\cal X}_\alpha )\psi _\alpha  (x)}\bigl\},\\
S_{\rm E,TE}[\varphi ] &=& \int \mbox{d}^{4} x\bigl\{{{\cal L} (\varphi(x))}\nonumber\\
&+& { \varphi (x)\sum\limits_{\alpha = 1}^2 \int \mbox{d}{\cal X}\delta ({\cal X} - {\cal X}_\alpha )\partial_n\psi _\alpha  (x)}\bigl\}.\hspace{.7cm}
\end{eqnarray}
\end{subequations}
By comparing  Eqs.~(\ref{Euclidean action}) and (\ref{expansion of generating functional for the interacting fields}), we can
rewrite Eqs.~(\ref{ZTMZTE}) as
\begin{subequations} \label{Euclidean action for interaction field}
\begin{eqnarray}
Z_{\rm TM} &=&\int \prod_{\alpha=1}^{2}{\cal {D}}[\psi_\alpha(x)]\nonumber\\&&
\times Z\left({\sum\limits_{\alpha  = 1}^2  \int {\mbox{d}{\cal X} \delta ({\cal X} -
{\cal X}_\alpha )} \psi _\alpha ({\cal X})}\right),\\
Z_{\rm TE} &=& \int \prod_{\alpha=1}^{2}{\cal {D}}[\psi_\alpha(x)]\nonumber\\&&
\times Z\left({\sum\limits_{\alpha  = 1}^2  \int {\mbox{d}{\cal X} \delta ({\cal X} - {\cal X}_\alpha )}
\partial_n\psi _\alpha ({\cal X})}\right),\;
\end{eqnarray}
\end{subequations}
where the $Z(\ldots)$ in the two integrands are the generating functionals of interacting fields
defined in Eq.~(\ref{expansion of generating functional for the interacting fields}) with imaginary time. From Eqs.~(\ref{Euclidean action for interaction field}) and (\ref{Green function for interaction field}) the respective partition functions can be
written as

\begin{equation}\label{effective partition functions}
Z_{\rm TM/TE} =\int \prod\limits_{\alpha  = 1}^2  {\cal {D}}[\psi
_\alpha (X_\alpha )]e^{- S_{\rm eff,{\rm TM/TE}} (\psi _\alpha
)},
\end{equation}
where the effective actions $S_{\rm eff,TM}$ and $S_{\rm eff,TE}$
are given by
\begin{subequations}
\begin{eqnarray}
&&S_{\rm eff,{TM}}(\psi _1 ,\psi _2 ) =\nonumber\\
&&\frac{1}{2}\sum
_{\alpha,\beta}\iint d{\cal X}_\alpha  d{\cal X}_\beta\,
\psi_\alpha({\cal X}_\alpha){\cal G}({\cal X}_\alpha,{\cal
X}_\beta)\psi_\beta({\cal X}_\beta),\\
&&S_{\rm eff,{TE}}(\psi _1 ,\psi _2 ) =\nonumber\\
&&\frac{1}{2}\sum
_{\alpha,\beta}\iint d{\cal X}_\alpha  d{\cal X}_\beta \,
\psi_\alpha({\cal X}_\alpha)[{\partial_{n_\alpha}}{\partial_{\rm
n_\beta}}{\cal G}({\cal X}_\alpha,{\cal X}_\beta)]\psi_\beta({\cal
X}_\beta).\nonumber\\
\end{eqnarray}
\end{subequations}
Here  the Green function of the fields after Wick
rotation is denoted by the new font ${\cal G}$. The partition functions defined by  (\ref{effective partition functions}) are calculated straightforwardly, and the results are given in Eqs.~(\ref{ZTETM}) and (\ref{GammaTETM}) in the main text.


\section{Green tensor for planar multilayer magnetodielectric media with gain}\label{Sec:Multilayer}
 For planar multilayer geometries as illustrated in Fig.~\ref{Fig:multilayer}, the electric permittivity and
magnetic permeability vary only in the $z$ direction, so we may
introduce a transverse spatial Fourier transform
as
\begin{eqnarray}
{{ \bfsfG}_{\rm EM}({\bf x}-{\bf x'},z,z',\imath\omega)}&=&\int
\mbox{d}^2{\bf q}\, e^{\imath {\bf
q}\cdot( {\bf x}-{\bf x'})}\bfsfG_{\rm EM}({\bf q},z,z',\imath\omega)\nonumber\\
\end{eqnarray}
where ${\bf q}$ is a vector parallel to the conductor. Toma{\v s}
uses this to arrive at the solution of Eq.~(\ref{differential equation for Green
function}) in lossy dielectric multilayers~\cite{Tomas1995a}. The generalization to
lossy magnetodielectric media can be found in Refs.~\cite{Chew1995a,Buhmann2005a}.
Here we briefly describe the results of a further nontrivial generalization, namely to magnetodielectrics with both loss and gain.
In our notation we follow Ref.~\cite{Tomas1995a}.

The Green tensor ${{\bfsfG}_{\rm EM}({\bf
q},z,z',\imath\omega)}$ assumes two different forms, depending on whether $z$ and $z'$ are located in the same layer or not. For $z'$ in
layer $j$ it is given by
\begin{widetext}
\begin{subequations}\label{Green tensor at imaginary frequencies}
\begin{eqnarray}
{{\bfsfG}_{\rm EM}({\bf
q},z,z',\imath\omega)}&=&\frac{1}{\varepsilon_0\varepsilon_j(\imath \omega
)\omega^2}\delta{(z-z')}\hat{z}\hat{z}+\frac{{\mu_0\mu _j (\imath \omega
)}}{{2 {\cal Q}_j }}\sum\limits_{\sigma  = \rm TE}^{\rm TM}
\xi_\sigma \frac{ { e^{ - {\cal Q}_j d_j } }}{{D_j^\sigma
}} \nonumber\\
&&[{\boldsymbol\varepsilon}^{\sigma\,>}_j ({\bf q},{\rm
i}\omega;z){\boldsymbol\varepsilon}^{\sigma\,<}_j (-{\bf
q},{\rm i}\omega;z')\Theta(z-z')+{\boldsymbol\varepsilon}^{\sigma\,<}_j ({\bf q},{\rm
i}\omega;z){\boldsymbol\varepsilon}^{\sigma\,>}_j (-{\bf q},{\rm
i}\omega;z')\Theta(z'-z) ],\,\,\,\,z\,\,\,\mbox{in layer}\,\,
j\nonumber\\ \\
{{\bfsfG}_{\rm EM}({\bf q},z,z',\imath\omega)}&=&\frac{{\mu_0\mu _l
(\imath \omega )}}{{2 {\cal Q}_l }}\sum\limits_{\sigma  = \rm
TE}^{\rm TM} \xi_\sigma \frac{ t_{l/j}^\sigma e^{ - ({\cal Q}_j d_j
+{\cal Q}_l d_l )}}{D_j^\sigma
} \nonumber\\
&\times&\left[{\frac{{\boldsymbol\varepsilon}^{\sigma\,>}_l ({\bf q},{\rm
i}\omega;z)}{D_{l/j}^{+,\sigma}}{\boldsymbol\varepsilon}^{\sigma\,<}_j
(-{\bf q},{\rm i}\omega;z')\Theta(l-j)+\frac{{\boldsymbol\varepsilon}^{\sigma\,<}_l ({\bf q},{\rm
i}\omega;z)}{D_{l/j}^{-,\sigma}}{\boldsymbol\varepsilon}^{\sigma\,>}_j
(-{\bf q},{\rm i}\omega;z')\Theta(j-l)} \right],\,\,z\,\,\, \mbox{in layer}\,
l\neq j \nonumber\\
\end{eqnarray}
\end{subequations}
\end{widetext}
where $\xi_{\rm TE}=-1$, $\xi_{\rm TM}=1$, and $\Theta(z)$ is the
usual unit step function and
\begin{subequations}
\begin{eqnarray}
{\boldsymbol\varepsilon}^{\sigma\,>}_j ({\bf q},{\rm i}\omega;z)&=&
{\bf e}_{\sigma j} ^+({\bf q})e^{ - {\cal Q}_j (z - d_j)}+r_{j^ +
}^\sigma{\bf e}_{\sigma j} ^-({\bf q})e^{  {\cal Q}_j (z -
d_j)},\nonumber\\ \\
{\boldsymbol\varepsilon}^{\sigma\,<}_j ({\bf q},{\rm i}\omega;z)&=&
{\bf e}_{\sigma j} ^-({\bf q})e^{  {\cal Q}_j z }+r_{j^
-}^\sigma{\bf e}_{\sigma j} ^+({\bf q})e^{ - {\cal Q}_j z
}.
\end{eqnarray}
\end{subequations}
Here $\sigma$ stands for ${\rm TE}$ or ${\rm TM}$, and
${\bf e}_{{\rm TE} j} ^{\pm}=(\hat{{\bf q}}\times \hat{z})_{j}$ and
${\bf
e}_{{\rm TM} j} ^{\pm}=\frac{-1}{q_j}({\rm i}|{\bf q}|\hat{z}\pm{\cal
Q}_j\hat{{\bf q}})_{j}$
are the polarization vectors for TE and TM polarized waves
propagating in the positive-/negative-$z$ direction, with $q_j \equiv \sqrt{\omega^2 \varepsilon_j({\rm i} \omega
)\mu_j({\rm i} \omega )/c^2}$  and
\begin{equation}\label{component of the wave vector}
{\cal Q}_j({\bf q},{\rm i} \omega ) = \sqrt {{q}^2+\omega^2
\varepsilon_j({\rm i} \omega )\mu_j({\rm i} \omega )/c^2 },
\end{equation}
which can be expressed in terms of the magnitude of the z-component $\kappa_j({\bf q},\omega )=\sqrt
{\omega^2 \varepsilon_j(\omega )\mu_j( \omega )/c^2 -{ q}^2}$  of the wave vector in layer $j$ as
${\cal Q}_j({\bf q},{\rm i}\omega ) = -{\rm i}\kappa_j({\rm i}\omega )$.

Here we arrive at a subtlety in  the determination of the Green tensor for active multilayer media:
the $z$-component of the wave vector, $\kappa_j({\bf q},\omega )$, is not always well defined
for real frequencies. The subtlety is that although the refractive index has no branch points in the
upper half-plane, $\kappa_j({\bf q},\omega )$ may have branch points there~\cite{Skaar2006a}. If there are such branch points, then $\kappa_j({\bf q},\omega )$
looses its usual physical interpretation. We will follow Refs.~\cite{Skaar2006b,Skaar2006a} and only consider active media without branch points where $\kappa_j(\omega )$ is meaningful for real frequencies. In that case
the signs of  $\mbox{Re}[\kappa_j(\omega )]$ and $\mbox{Im}[\kappa_j(\omega )]$ are identical to
those of $\mbox{Re}[n_j(\omega )]$ and $\mbox{Im}[n_j(\omega )]$, respectively, where $n_j$ is refractive index of $j$-th layer (see Refs.~\cite{Skaar2006b,Skaar2006a}).
Other quantities in Eqs.~(\ref{Green tensor at imaginary frequencies}) that still need to be defined are
\begin{subequations}
\begin{eqnarray}
D_j^\sigma  & = & 1 - r_{j^ -  }^\sigma  r_{j^ + }^\sigma e^{ - 2Q_j
d_j }, \\
D_{l/j}^{\pm ,\sigma} & = & 1 - r_{l\pm  }^{ \sigma}  r_{ll\mp1/j
}^\sigma e^{ - 2Q_l d_l }
\end{eqnarray}
\end{subequations}
where  $r_{j^ - }^\sigma$ and $r_{j^ + }^\sigma$ are the generalized
coefficients for reflection at the left/right boundary of layer $j$,
which can be calculated with the aid of the recursive relations
\cite{Chew1995a,Buhmann2005a}
\begin{subequations}\label{recursive relations}
\begin{equation}
r_{j\, \pm }^{\rm TE}  = \frac{{\left( {\frac{{\mu _{j \pm 1}
}}{{{\cal Q}_{j \pm 1} }} - \frac{{\mu _j }}{{{\cal Q}_j }}} \right)
+ \left( {\frac{{\mu _{j \pm 1} }}{{{\cal Q}_{j \pm 1} }} +
\frac{{\mu _j }}{{{\cal Q}_j }}} \right)e^{ - 2{\cal Q}_{j \pm 1}
d_{j \pm 1} } r_{j \pm 1 \pm }^{\rm TE} }}{{\left( {\frac{{\mu _{j
\pm 1} }}{{{\cal Q}_{j \pm 1} }} + \frac{{\mu _j }}{{{\cal Q}_j }}}
\right) + \left( {\frac{{\mu _{j \pm 1} }}{{{\cal Q}_{j \pm 1} }} -
\frac{{\mu _j }}{{{\cal Q}_j }}} \right)e^{ - 2{\cal Q}_{j \pm 1}
d_{j \pm 1} } r_{j \pm 1 \pm }^{\rm TE} }}
\end{equation}
for TE-polarized light, and for TM polarization
\begin{equation}
r_{j\, \pm }^{\rm TM}  = \frac{{\left( {\frac{{\varepsilon _{j \pm 1}
}}{{{\cal Q}_{j \pm 1} }} - \frac{{\varepsilon _j }}{{{\cal Q}_j }}}
\right) + \left( {\frac{{\varepsilon _{j \pm 1} }}{{{\cal Q}_{j \pm 1}
}} + \frac{{\varepsilon _j }}{{{\cal Q}_j }}} \right)e^{ - 2{\cal Q}_{j
\pm 1} d_{j \pm 1} } r_{j \pm 1 \pm }^{\rm TM} }}{{\left(
{\frac{{\varepsilon _{j \pm 1} }}{{{\cal Q}_{j \pm 1} }} +
\frac{{\varepsilon _j }}{{{\cal Q}_j }}} \right) + \left(
{\frac{{\varepsilon _{j \pm 1} }}{{{\cal Q}_{j \pm 1} }} -
\frac{{\varepsilon _j }}{{{\cal Q}_j }}} \right)e^{ - 2{\cal Q}_{j \pm
1} d_{j \pm 1} } r_{j \pm 1 \pm }^{\rm TM} }}.
\end{equation}
\end{subequations}
For a finite number of layers there is only a finite number of relations to be solved,
since for the leftmost and rightmost layers one should take $r^{\rm TE,TM}_{1-}=0$, $r^{\rm
TE,TM}_{n+}=0$, and $d_1=d_n=0$. From the definition of the Fresnel
coefficients introduced it follows that they satisfy
\begin{equation}\label{transmision recursive relations}
r_{i/j/k }^{\sigma}=\frac{1}{D_j^\sigma}[r_{i/j }^{\sigma}+(t_{i/j
}^{\sigma}t_{j/i }^{\sigma}-r_{i/j }^{\sigma}r_{j/i
}^{\sigma})r_{j/k }^{\sigma}e^{-2{\cal Q}_jd_j}],
\end{equation}
in the notation of Ref.~\cite{Tomas1995a}.

We have hereby specified the rather complicated expression for the Green tensor $\bfsfG$ of Eq.~(\ref{Green
tensor at imaginary frequencies}), and we still need to relate it to the
Green tensors ${\cal G}^{\rm TE,TM}({\bf q},z,z',\imath\omega)$ in terms of which the Casimir force is expressed in Eq.~(\ref{GammaTETM}).
By using the ordinary coordinates according
to the convention of Schwinger {\em et al.}~\cite{Schwinger1978a}
and choosing $(\hat{{q}},\hat{z},\hat{{q}}\times\hat{z})\longrightarrow(\hat{x},\hat{z},-\hat{y})$
\cite{Ellingsen2007a}, the  transverse electric and transverse
magnetic Green functions satisfy
\begin{subequations}\label{GTETMmultilayer}
\begin{eqnarray}
&&[ - \frac{\partial }{{\partial z}}\frac{1}{{\mu
({\rm i}\omega,z)}}\frac{\partial }{{\partial z}} + \frac{{{q}^2
}}{{\mu ({\rm i}\omega,z)}} + \frac{\omega ^2 \varepsilon
({\rm i}\omega,z)}{c^2}]{\cal G}^{\rm TE}({\bf q},z,z',{\rm i}\omega),\nonumber\\
&&=\mu_0
\delta (z - z'),\label{50/6}\\
&&[- \frac{\partial }{{\partial z}}\frac{1}{{\varepsilon
({\rm i}\omega,z)}}\frac{\partial }{{\partial z}} + \frac{{{q}^2
}}{{\varepsilon ({\rm i}\omega,z)}} + \frac{\omega ^2 \mu ({\rm i
}\omega,z)}{c^2}]{\cal G}^{\rm TM}({\bf q},z,z',{\rm i}\omega) \nonumber\\
&&= \mu_0 \delta(z - z').
\end{eqnarray}
\end{subequations}
We checked with some lengthy but straightforward calculations that these
Green functions ${\cal G}^{\rm TE,TM}({\bf q},z,z',{\rm i}\omega)$ with $z'$ and $z$ in layers $j$ and $l$
respectively, can be written very elegantly in terms of $\bfsfG({\bf q},z,z',\imath\omega)$ as
\begin{subequations}\label{scaler Green function}
\begin{eqnarray}
&&\bfsfG_{yy}({\bf q},z,z',{\rm i}\omega)={\cal G}^{\rm TE}({\bf
q},z,z',{\rm i}\omega),\\
&&\bfsfG_{zz}({\bf q},z,z',{\rm i}\omega)=\delta_{lj}\frac{\delta (z
- z')}{\varepsilon_0\varepsilon({\rm i}\omega,z)\omega^2}\nonumber\\
&&+\frac{{ q}^2 c^2}{\varepsilon({\rm i}\omega,z)\varepsilon({\rm
i}\omega,z')\omega^2}{\cal G}^{\rm TM}({\bf q},z,z',{\rm i}\omega)
\end{eqnarray}
\end{subequations}
It is important to point out what has been achieved here: by the identifications~(\ref{scaler Green function}) we have found solutions for the scalar Green functions
${\cal G}^{\rm TE,TM}$ that are defined by the equations~(\ref{GTETMmultilayer}), with boundary conditions that follow from the continuity of $H_x$, $H_y$, and $\mu H_z$, and of $E_x$,
$E_y$, and $\varepsilon E_z$.


\end{document}